\documentclass[preprint,onecolumn,times]{elsarticle}

\usepackage[T1]{fontenc}
\usepackage[utf8]{inputenc}
\usepackage[english]{babel}

\usepackage{amsmath}
\usepackage{newtxtext,newtxmath}

\usepackage[version=4]{mhchem}
\usepackage{siunitx}

\usepackage{graphicx}
\usepackage{subcaption}
\usepackage[export]{adjustbox}

\usepackage{float}
\usepackage{placeins}

\usepackage{xspace}
\usepackage{xcolor}

\usepackage{hyperref}
\hypersetup{
  colorlinks=true,
  citecolor=blue,
  urlcolor=blue,
  linkcolor=blue
}

\journal{Publication}

\begin{document}
\begin{frontmatter}

\title{Fluctuation effect on Nonlinear Transport and Nernst--Ettingshausen Response in Two-Dimensional Superconductors under electric and magnetic field}

\author[label1]{Tran Ky Vi}
\author[label2]{Bui Duc Tinh\corref{cor1}}
\author[label2]{Ngo Quang Duc}
\author[label2]{Chu Gia Bao}
\author[label6]{Le Viet Hoang}
\author[label7,label8]{Le Xuan The Tai}
\author[label1]{Nguyen Viet Hung}


\affiliation[label1]{
  organization={School of Materials Science and Engineering (SMSE), Hanoi University of Science and Technology (HUST)},
  addressline={No. 1 Dai Co Viet Str.},
  city={Hanoi},
  country={Vietnam}
}

\affiliation[label2]{
  organization={Faculty of Physics, Hanoi National University of Education},
  addressline={136 Xuan Thuy Str.},
  city={Hanoi},
  country={Vietnam}
}

\affiliation[label6]{
  organization={University of Science and Technology of Hanoi (USTH)},
  addressline={18 Hoang Quoc Viet Str.},
  city={Hanoi},
  country={Vietnam}
}

\affiliation[label7]{
  organization={Atomic Molecular and Optical Physics Research Group, Science and Technology Advanced Institute, Van Lang University},
  city={Ho Chi Minh City},
  country={Vietnam}
}

\affiliation[label8]{
  organization={Faculty of Applied Technology, Van Lang School of Technology, Van Lang University},
  city={Ho Chi Minh City},
  country={Vietnam}
}

\cortext[cor1]{Corresponding author. Email: tinhbd@hnue.edu.vn}

\begin{abstract}
IN this paper, we present a unified theoretical study of fluctuation-dominated transport and transverse thermoelectric response in two-dimensional superconducting films subjected to out-of-plane magnetic fields and electric-field drive. 
Our approach is based on the time-dependent Ginzburg--Landau equation with Langevin thermal noise, in which interaction effects of fluctuating Cooper pairs are incorporated self-consistently at the Gaussian (Hartree) level. 
We derive closed-form expressions for the fluctuation-induced Cooper-pair density, the renormalized resistance $R(T,B_\perp)$, and the nonlinear current response $J(E,B_\perp)$, explicitly accounting for the feedback of the electric field on the fluctuation spectrum. 
A central result is the emergence of an intrinsic S-shaped nonlinear $J$--$E$ (or $I$--$V$) characteristic, featuring a negative-differential segment and multivalued solutions under voltage control. 
Within this framework, we introduce a physically transparent procedure to identify characteristic instability scales, such as the magnetic field $B^{\ast}$ (or equivalently $B_{\chi}$), which marks the terminal point of the S-shaped instability where the nonlinear response becomes single-valued. 
In parallel, we analyze the off-diagonal Peltier coefficient $\alpha_{xy}$ as a direct probe of the transverse thermoelectric response of superconducting fluctuations. 
The theory is validated through systematic comparisons with recent experimental measurements of multi-field $R(T)$ curves, nonlinear $I$--$V$ characteristics, and $\alpha_{xy}$ data across a broad range of thin-film superconducting materials.
\end{abstract}

\end{frontmatter}


\section{Introduction}
\label{sec:introduction}

Since its discovery by Heike Kamerlingh Onnes in 1911~\cite{onnes1911resistance}, superconductivity has stood as one of the paradigmatic manifestations of macroscopic quantum coherence. Beyond its foundational importance, superconductivity has continuously motivated both fundamental and applied research, ranging from magnetic levitation and superconducting magnets to medical imaging and quantum sensing technologies~\cite{pollard1984maglev,damadian1971tumor,sternickel2006biomagnetism}. More than a century after its discovery, the field remains vibrant, driven in large part by the emergence of low-dimensional and strongly fluctuating superconductors.

A decisive shift in perspective occurred with the realization that superconductivity in reduced dimensionality is governed not only by coherent condensate physics, but also—often predominantly—by thermal and dynamical fluctuations of the order parameter. In two-dimensional (2D) systems, reduced phase space and enhanced thermal agitation qualitatively modify the superconducting transition: instead of a sharp mean-field onset, experiments reveal broad crossover regimes in which short-lived Cooper pairs and vortex-like excitations persist over extended temperature and magnetic-field ranges. This fluctuation-dominated phenomenology has been systematically developed over several decades and synthesized in the fluctuation theory of superconductivity formulated by Larkin and Varlamov~\cite{larkin2005theory,PhysRevB.44.7078,PhysRevLett.65.259,PhysRevB.48.12951}, where superconducting fluctuations generate excess conductivity, diamagnetism, and entropy transport far beyond the immediate vicinity of the critical temperature.

One of the most robust experimental manifestations of this physics is the broadening of the resistive transition under a perpendicular magnetic field. Rather than collapsing abruptly to zero resistance, the $R(T,B_\perp)$ curves of 2D superconductors exhibit smooth, field-dependent crossovers extending deep into the nominally normal state~\cite{larkin2005theory}. This behavior, observed across a wide variety of materials—including cuprates, disordered thin films, transition-metal dichalcogenides, and iron-based superconductors—reflects the gradual suppression of fluctuation correlations by magnetic confinement and cannot be accounted for within static or purely microscopic descriptions, nor can it be attributed exclusively to strong electronic correlation effects treated at the equilibrium level. 

While linear-response transport has long served as the primary tool for establishing fluctuation-induced conductivity and magnetization, its scope is inherently limited: by construction, it probes only near-equilibrium dynamics and is largely insensitive to nonlinear feedback between superconducting correlations and external drive. A qualitatively different and far more discriminating window opens in the nonlinear regime, where an applied electric field drives the system out of equilibrium. In this context, S-shaped current--voltage ($I$--$V$) or current--electric-field ($J$--$E$) characteristics have emerged as one of the most distinctive fingerprints of driven superconductivity. An S-shaped curve, defined by the presence of a negative-differential segment ($dI/dV<0$ or $dJ/dE<0$), implies multivalued transport solutions and therefore signals an intrinsic dynamical instability of the superconducting state under drive. Such behavior is fundamentally nontrivial: it is absent in ordinary conductors and directly reflects feedback between superconducting correlations and nonequilibrium forcing.

Historically, however, the experimental identification of S-shaped transport has been severely constrained by stability considerations. In current-biased measurements, the negative-differential branch is generically unstable and is bypassed by abrupt voltage switching, phase-slip formation, or filamentary breakdown. As a result, despite more than a century of superconductivity research since Onnes’ discovery of vanishing resistance in 1911~\cite{onnes1911resistance}, clear and continuous S-shaped $I$--$V$ curves have remained rare, repeatedly giving rise to debates over whether reported nonlinearities originate from intrinsic superconducting physics or from extrinsic effects such as heating, inhomogeneity, or contact artifacts.

The modern theoretical history of S-shaped transport begins with the seminal work of Vodolazov \textit{et al.}~\cite{vodolazov2003current}, who demonstrated within the time-dependent Ginzburg--Landau (TDGL) framework~\cite{schmid1966time} that quasi-one-dimensional superconductors admit multivalued solutions in the constant-voltage regime, yielding an ``unusual S behavior'' as an intrinsic superconducting response rather than a measurement artifact. Subsequent nanowire experiments~\cite{michotte2004development} clarified an important practical limitation of this prediction: under current bias, the unstable branch is typically hidden by switching and phase-slip dynamics, explaining why continuous S-shaped curves are seldom resolved in standard transport measurements.

As nonlinear transport studies expanded to correlated and higher-dimensional superconductors, additional limitations of early interpretations became apparent. Experiments on cuprates~\cite{wei2005nanoscale,kaminski2016destroying} showed that strong electrical drive can suppress phase coherence while leaving the pairing gap largely intact, indicating that nonequilibrium superconducting dynamics and strong fluctuations, rather than simple Joule heating, govern the nonlinear response \cite{kwak2023fluctuation}. Nevertheless, a unifying microscopic mechanism capable of generating S-shaped behavior in higher dimensions—particularly in the absence of well-defined phase-slip centers or ordered vortex motion—remained elusive.

A decisive conceptual advance was achieved by Qiao \textit{et al.}~\cite{qiao2018dynamical}, who demonstrated that S-shaped transport can arise intrinsically from superconducting fluctuations within TDGL theory supplemented by a self-consistent Gaussian (Hartree) treatment of thermal Cooper-pair interactions~\cite{larkin2005theory, schmid1969diamagnetic,ullah1991effect,rosenstein2010ginzburg,kovner1989covariant,rosenstein1989covariant}. In this framework, the electric field feeds back onto the fluctuation spectrum itself, renormalizing the effective Ginzburg--Landau mass and rendering the superconducting contribution to the current nonmonotonic in $E$, even at $B=0$ and without invoking vortices. The instability is predicted quantitatively through the simultaneous conditions $dJ/dE=0$ and $d^{2}J/dE^{2}=0$~\cite{qiao2018dynamical}. Subsequent extensions~\cite{dang2022voltage,vi2024fluctuation} established that this fluctuation-driven S-shaped instability persists under magnetic field and is continuously suppressed as superconducting correlations are weakened, rather than being generated by the field itself.

From a historical perspective, a key limitation remained experimental rather than theoretical: resolving a continuous S-shaped branch requires genuine voltage-controlled stabilization of an intrinsically unstable nonlinear state. This obstacle was only overcome very recently, when Nagahama \textit{et al.}~\cite{nagahama2025two} implemented carefully stabilized voltage scans under magnetic field in a genuinely two-dimensional $\mathrm{Bi}_{2}\mathrm{Te}_{3}/\mathrm{Fe(Se,Te)}$ heterostructure. Their 2025 \emph{Physical Review Letters} experiment directly resolved the continuous S-shaped branch that had long been predicted but systematically hidden in current-biased measurements. In this sense, the experiment should be viewed not as the discovery of a new phenomenon, but as the long-awaited experimental realization of an intrinsically multivalued nonequilibrium superconducting response whose existence had been constrained for decades by measurement-induced instabilities.

Taken together, the historical development of S-shaped transport highlights a recurring theme: theoretical predictions of intrinsic multivalued nonlinear response appeared early, but experimental access was delayed by fundamental stability limitations. Only with the convergence of fluctuation-based theory and genuinely voltage-controlled measurement protocols has the S-shaped instability emerged as a quantitative and unambiguous probe of nonequilibrium superconducting dynamics.

Fluctuations in two-dimensional superconductors are not limited to charge transport; they also carry entropy and energy, giving rise to pronounced transverse thermoelectric responses. The off-diagonal Peltier coefficient $\alpha_{xy}$ plays a central role in this context, governing both the Nernst effect and its Onsager-reciprocal Ettingshausen effect \cite{PhysRevLett.65.2066, PhysRevB.83.020506, PhysRevLett.126.077001}. Unlike longitudinal transport coefficients, $\alpha_{xy}$ directly measures transverse entropy flow per unit charge and is therefore particularly sensitive to fluctuating Cooper pairs and vortex-like excitations. Measurements of $\alpha_{xy}$ have provided compelling thermodynamic evidence for superconducting fluctuations persisting well above $T_c$ in cuprates, disordered films, and iron-based systems \cite{RevModPhys.90.015009,PhysRevResearch.4.013186}.

Iron-based superconductors constitute a particularly timely and important platform for this unified fluctuation perspective. Their short coherence lengths, multiband electronic structure, and proximity to competing electronic orders naturally enhance fluctuation effects, placing them on equal conceptual footing with cuprates and strongly disordered thin films. The ability to describe such diverse material classes within a single mesoscopic framework underscores the universality of fluctuation-based approaches beyond microscopic pairing mechanisms. Recent developments in fluctuation spectroscopy and nonequilibrium superconducting transport further reinforce this universality, as highlighted by comprehensive theoretical treatments of fluctuation phenomena~\cite{RevModPhys.90.015009}, nonlinear transverse thermoelectric responses~\cite{PhysRevResearch.4.013186}, and voltage-controlled two-dimensional superconducting devices~\cite{nagahama2025two}.

In this work, we present a unified treatment of fluctuation-driven transport and thermodynamics in two-dimensional superconducting films. Our approach is based on the time-dependent Ginzburg--Landau equation (TDGLE) with Langevin thermal noise, where the quartic interaction is incorporated within a self-consistent fluctuation (Hartree) approximation. Within this framework, we compute the renormalized resistance $R(T,B_\perp)$, the full nonlinear current response $J(E,B_\perp)$ including intrinsic S-shaped instabilities, and the transverse thermoelectric coefficient $\alpha_{xy}$ associated with the Nernst and Onsager-reciprocal Ettingshausen effects, and we systematically compare these theoretical results with recent experimental measurements.


The paper is organized as follows. 
Section~\ref{sec:2} introduces the quasi-two-dimensional time-dependent Ginzburg--Landau (TDGL) model with Langevin thermal noise and establishes the dimensionless variables and characteristic scales used throughout. 
Section~3 develops the self-consistent Gaussian (Hartree) approximation, derives the Green's function of the linearized TDGL operator, and obtains closed expressions for the fluctuation Cooper-pair density, the renormalized resistance, and the heat current together with the transverse thermoelectric coefficient $\alpha_{xy}$. 
Section~4 presents the numerical results and systematic comparisons with experiments, including fluctuation-controlled resistive broadening under perpendicular magnetic fields, voltage-controlled nonlinear transport with S-shaped $I$--$V$ characteristics and associated instability boundaries, and the transverse thermoelectric response quantified by $\alpha_{xy}$. 
Section~5 discusses the physical implications and the scope of the fluctuation-based mesoscopic description across different material platforms. 
Finally, Section~6 summarizes the main conclusions and outlines perspectives for nonequilibrium superconductivity in two-dimensional and iron-based systems.

\section{The quasi two-dimensional TDGL model with Langevin thermal noise}
\label{sec:2}

\subsection{The quasi two-dimensional description}

We consider an effectively two-dimensional superconducting system, appropriate for ultrathin films or atomically thin materials, subjected to a magnetic field applied perpendicular to the sample plane. In this regime, the superconducting order parameter is assumed to be uniform across the film thickness, allowing the dynamics to be described by a two-dimensional time-dependent Ginzburg--Landau (TDGL) equation with Langevin thermal noise.

The perpendicular magnetic field $ \mathbf{B}=B\hat{z} $ induces vortex-like excitations confined within the plane, while an in-plane electric field $ \mathbf{E} $ drives dissipative transport. This effective two-dimensional formulation captures the essential fluctuation physics governing transport and thermodynamic responses near the superconducting transition.

\begin{figure}[htbp]
  \centering
  \includegraphics[width=0.8\linewidth]{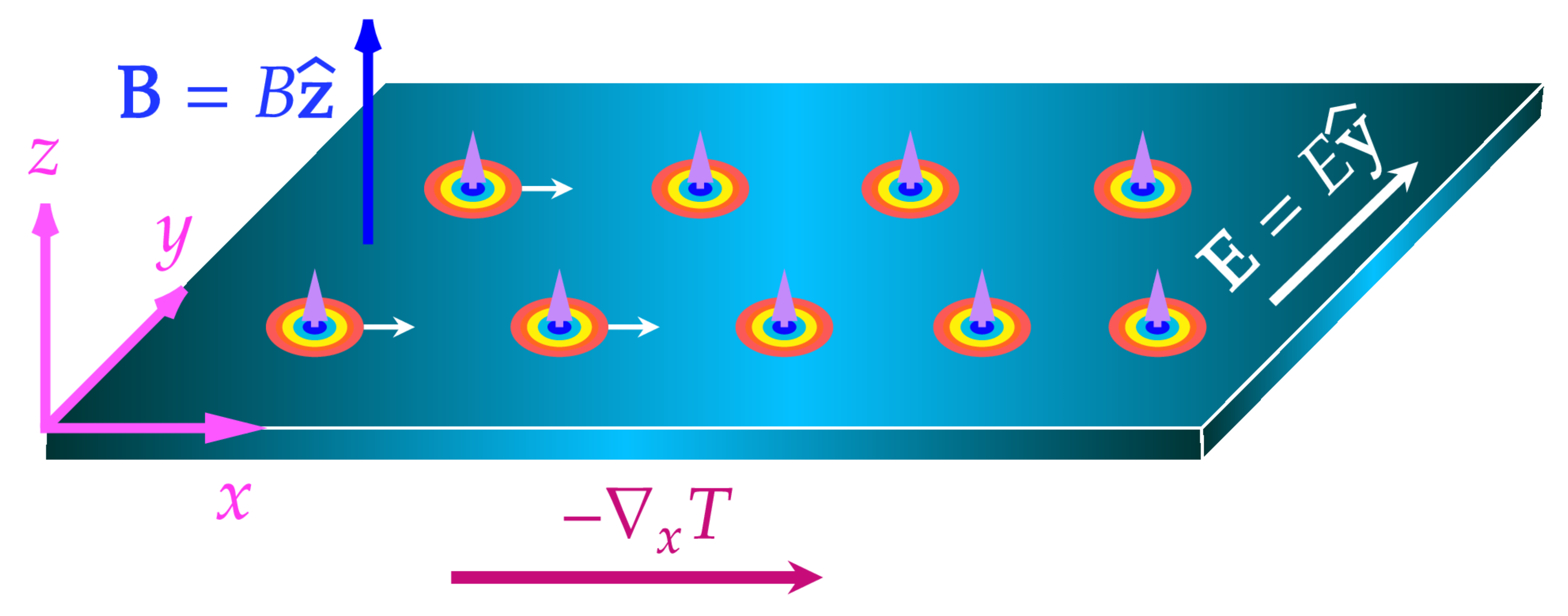}
\caption{Schematic of an effectively two-dimensional superconducting film sample in an out-of-plane magnetic field $\mathbf{B}=B\hat{z}$, with in-plane electric field $\mathbf{E}$ driving transport and vortex-like excitations confined to the $xy$ plane.}

  \label{fig:diagram-20250902}
\end{figure}

In two spatial
dimensions, the order parameter $\psi(\mathbf r,t)$, which decribes the dynamics of Cooper pairs, couples  to
the scalar and vector potentials, $\Phi(\mathbf r,t)$ and
$\mathbf A(\mathbf r,t)$, via the gauge-covariant derivatives
$\mathcal D_t \equiv \frac{\partial}{\partial t} + \frac{\text{i} e_s}{\hbar}\,\Phi,
\qquad
\boldsymbol{\mathcal D} \equiv \nabla - \frac{\text{i} e_s}{\hbar}\,\mathbf A,
$ where $e_s=-2|e|$ and $m_s$ are the effective charge and mass of a Cooper
pair. A convenient free energy density yielding the TDGLE and Maxwell
equations is
\begin{equation}\label{eq:TDGL-L}
\mathcal{F}=\frac{\hbar^{2}}{2m_s}\bigl|\boldsymbol{\mathcal{D}}\Psi\bigr|^{2}
+\alpha\,(T-T_c)|\Psi|^{2}
+\frac{\beta}{2}|\Psi|^{4}
+\frac{\mathbf{B}}{2\mu_{0}},
\end{equation}
where $\mathbf B=\nabla\times\mathbf A$ is the magnetic induction corresponding with applied vector potential field $\textbf{A}$. The first term penalizes spatial
variations through the covariant gradient; the quadratic and quartic
terms provide the condensate free-energy landscape; and the final term
captures the magnetic energy, the field-energy contribution reduces
accordingly, a case frequently adopted in thin-film (2D) analyses of
vortex dynamics and relaxation phenomena. The totally free energy is 

\begin{equation}\label{eq:TDGL-L1}
F_{GL} =L\int \mathrm{d}^2\mathbf{r}\mathcal{F},
\end{equation}
where, $L$ denotes the thickness of the thin film, $m_s$ is the effective mass of the Cooper pair, and $T_{c0}$ represents the mean field critical temperature. The superconducting coherence length is defined as $\xi^2 = \frac{\hbar^2}{2m_s \alpha T_{c0}}$ while the magnetic penetration depth is given by $\lambda^2
=
\frac{\beta m_s}{4\mu_0 e_s^{\,2}\alpha T_{c0}},
$ where $\alpha$ and $\beta$ are the Ginzburg-Landau coefficients. When external electric and magnetic fields are applied, the relaxation dynamics of the superconducting order parameter are governed by the gauge-invariant TDGLE. In the presence of uniform electric and out-of-plane magnetic fields, we work in the gauge
$\Phi(\mathbf{r}) = -Ey$, and $\mathbf{A}(\mathbf{r}) = -By\,\hat{\mathbf{e}}_x ,
$
so that the gauge-invariant TDGL equation consistently incorporates both fields.
The TDGLE including Langevin white noise can be written as \cite{abrikosov2017fundamentals}
\begin{equation}
\frac{\hbar^{2}\gamma}{2m_s}\,\mathcal D_t \Psi
=
-\frac{1}{L}\frac{\delta F_{GL}}{\delta \Psi^{\ast}}
+\zeta ,
\label{TDGL_I}
\end{equation}
where \(F_{GL}\) is the Ginzburg--Landau free energy functional and \(\zeta\) represents thermal noise.

According to the fluctuation--dissipation theorem, the thermal forces that induce thermodynamical fluctuations satisfy
\begin{equation}
\bigl\langle
\zeta^{\ast}(\mathbf{r},\tau)\,
\zeta(\mathbf{r}^{\prime},\tau^{\prime})
\bigr\rangle
=
\frac{\hbar^{2}\gamma T}{m_s L}\,
\delta(\mathbf{r}-\mathbf{r}^{\prime})\,
\delta(\tau-\tau^{\prime}),
\end{equation}
where \( \langle\cdots\rangle \) denotes a thermal average
\cite{larkin2005theory,halperin2019theory}.

The electric current density consists of two components,
\(
\mathbf{J}^{e}=\mathbf{J}_{n}^{e}+\mathbf{J}_{s}^{e},
\)
where the normal contribution
\(
\mathbf{J}_{n}^{e}=\sigma_{n}\mathbf{E}
\)
describes Ohmic dissipation. The superconducting contribution \( \mathbf{J}_{s}^{e} \) is obtained by performing a functional variation of the Ginzburg--Landau free energy in Eq.~(\ref{eq:TDGL-L1}) with respect to the vector potential \( \mathbf{A} \). The electric supercurrent density assumes the standard gauge--invariant form
\begin{equation}
\mathbf{J}_s^e=-\frac{\mathrm{i} e_s \hbar}{m_s}\left(\Psi^* \nabla \Psi-\Psi \nabla \Psi^*\right)-\frac{e_s^2}{m_s}|\Psi|^2 \mathbf{A} .
\end{equation}

Within the TDGL framework, heat transport associated with the dynamics of the superconducting order parameter is described by the heat current density \( \mathbf{J}^{h} \). This quantity follows directly from the microscopic energy current obtained via Noether's theorem associated with time--translation invariance of the TDGL Lagrangian. After subtracting the electrochemical work contribution \( \Phi\,\mathbf{J}^{e} \), the intrinsic heat current is given by the manifestly gauge--invariant expression
\begin{equation}
\mathbf{J}^{h}
=
-\frac{\hbar^{2}}{2m_s}
\Big[
(\mathcal D_t\Psi)^{\ast}\,\boldsymbol{\mathcal D}\Psi
+
(\mathcal D_t\Psi)\,(\boldsymbol{\mathcal D}\Psi)^{\ast}
\Big].
\end{equation}
This expression is obtained without additional assumptions or linear--response approximations and represents the mesoscopic heat current of the TDGL theory.

In linear response, the transport electric and heat currents are related to the applied electric field and temperature gradient through the Onsager matrix,
\(
\displaystyle\binom{\mathbf{J}^{e}}{\mathbf{J}^{h}_{\mathrm{tr}}}
=
\begin{pmatrix}
\sigma & \alpha \\
T\alpha & \kappa
\end{pmatrix}
\displaystyle\binom{\mathbf{E}}{-\nabla T},
\)
where the off--diagonal coefficients obey the Onsager reciprocity relation \(\tilde{\alpha}=T\alpha\).
The transverse component \(\alpha_{xy}\) therefore has a dual interpretation: it characterizes either the transverse electric current generated by a longitudinal temperature gradient (Nernst effect), or equivalently the transverse heat current induced by an electric field (Ettingshausen effect). In the isothermal configuration considered throughout this work, where the
temperature gradient vanishes ($\nabla T = 0$), the transverse transport heat
current induced by an applied electric field reduces to
\begin{equation}
J^{h}_{x,\mathrm{tr}}
=
\tilde{\alpha}_{xy} E_y
=
T\,\alpha_{xy} E_y ,
\label{eq:heat_current_isothermal}
\end{equation}
where $\tilde{\alpha}_{xy}=T\alpha_{xy}$ follows from the Onsager reciprocity
relation between thermoelectric coefficients.

Consequently, within linear-response theory, the Ettingshausen coefficient
$\alpha_{xy}$ can be directly extracted from the mesoscopic heat current as
\begin{equation}
\alpha_{xy}
=
\frac{\langle J_x^{h} \rangle}{T E_y},
\label{eq:alpha_xy_definition}
\end{equation}
providing a direct link between microscopic energy transport and experimentally
accessible transverse thermoelectric measurements
(see Ref.~\cite{PhysRevLett.89.287001}).
Accordingly, after evaluating the fluctuation-induced heat current within the TDGL framework, one can directly infer the transverse thermoelectric coefficient and thus connect microscopic energy transport to the experimentally measured Nernst response.

\subsection{Dimensionless variables and characteristic scales}

To facilitate the numerical analysis and identify the fundamental physical regimes, we introduce a set of dimensionless variables by scaling the physical quantities with respect to their characteristic units. In this framework, spatial coordinates are normalized by the coherence length $\xi$, while time is measured in units of the Ginzburg-Landau relaxation time, defined as $\tau_{GL} = \gamma\xi^{2}/2$. The order parameter is scaled by its equilibrium value as $\Psi = \psi_0 \psi$, where $\psi_0 = \sqrt{2\alpha T_{c0}/\beta}$. 
Furthermore, the electric field is expressed in terms of the characteristic scale
$E_{GL} = 2\hbar / (\gamma |e_s| \xi^{3})$ such that $E = E_{GL}\mathcal{E}$.
The perpendicular magnetic field is rendered dimensionless as
$\mathcal{B}=B/\bar{B}_{c2}$, where $\bar{B}_{c2}$ denotes the characteristic (mean-field) upper critical field scale.
By introducing these dimensionless variables, the TDGLE in Eq.~(\ref{TDGL_I}) can be cast into the normalized form that governs the dynamics of the superconducting order parameter,
\begin{equation}
\overline{\mathcal{D}}_\tau \psi
-\frac{1}{2}\,\overline{\boldsymbol{\mathcal{D}}}^{\,2}\psi
-\frac{1-\mathcal{T}}{2}\,\psi
+|\psi|^{2}\psi
=\overline{\zeta},
\label{eq:TDGL_dimensionless}
\end{equation}
where $\mathcal{T}=T/T_{c0}$ is the reduced temperature,
$\overline{\mathcal{D}}_\tau = \partial_\tau + \mathrm{i}\mathcal{E}y$
is the gauge-covariant time derivative, and
$\overline{\boldsymbol{\mathcal{D}}}^{\,2}
= (\partial_x - \mathrm{i}\mathcal{B}y)^2 + \partial_y^2$
denotes the dimensionless Laplacian incorporating the effect of the perpendicular magnetic field.
. 

The stochastic thermal noise $\overline{\zeta}$ is assumed to be Gaussian and white-noise distributed, satisfying the fluctuation-dissipation relation:\begin{equation}\langle \overline{\zeta}^{\ast}(\mathbf{r},\tau) \overline{\zeta}(\mathbf{r}',\tau') \rangle = 2\omega \mathcal{T} \delta(\mathbf{r}-\mathbf{r}')\delta(\tau-\tau').\end{equation}

The fluctuation strength is characterized by the parameter
$\omega\simeq\pi\sqrt{2\,\mathrm{Gi}_{2D}}$,
where the two-dimensional Ginzburg number is defined as
\begin{equation}
\mathrm{Gi}_{2D}
=
\frac{1}{2}
\left(
\frac{2\mu_{0}\,e_s^{2}\,\lambda^{2}\,T_{c0}\,\gamma}
{\hbar^{2}\,\xi}
\right)^{2}.
\end{equation}

Physical observables are related to their dimensionless representations through the characteristic current scales \(J_{GL}^{e}\) and \(J_{GL}^{h}\).  
In SI units, the natural scale for the electric current density follows from the equilibrium condensate amplitude
\(\Psi_0^{2}=2\alpha T_{c0}/\beta\).
Combining the microscopic parameters of the Ginzburg--Landau theory, this scale can be written in several equivalent forms,
\begin{equation}
J_{GL}^{e}
=
\frac{|e_s|\,\hbar}{m_s\,\xi}\,\Psi_0^{2}
=
\frac{\bar{B}_{c2}\,\xi}{2\mu_{0}\,\lambda^{2}}
=
\frac{\Phi_{0}}{4\pi\mu_{0}\,\lambda^{2}\,\xi},
\end{equation}
where the upper critical field and the flux quantum are defined as
\({B}_{c2}=\Phi_{0}/(2\pi\xi^{2})\) and \(\Phi_{0}=h/|e_s|\), respectively.

In an analogous manner, the characteristic scale for the heat current density is introduced as

\begin{equation}
J_{GL}^{h}
=
\frac{\hbar\, \bar{B}_{c2}}{\mu_0\,\gamma_s\,|e_s|\,\kappa^{2}\,\xi^{3}}
=
\frac{\hbar\,\Phi_0}{2\pi\,\mu_0\,\gamma_s\,|e_s|\,\lambda^{2}\,\xi^{3}} .
\label{eq:JGL_h}
\end{equation}

The physical electric and heat current densities, $\mathbf{J}^{e}$ and
$\mathbf{J}^{h}$, are normalized by the characteristic scales
$J_{GL}^{e}$ and $J_{GL}^{h}$ to define the corresponding dimensionless
quantities. The dimensionless electric current density then reads
\begin{equation}
\mathbf{j}^{e}
=
-\frac{\mathrm{i}}{2}
\left(
\psi^{*}\,\bar{\nabla}\psi
-
\psi\,\bar{\nabla}\psi^{*}
\right)
-
|\psi|^{2}\,\mathbf{a},
\label{eq:je_dimensionless}
\end{equation}
where the dimensionless vector potential is given by
\(\mathbf{a}\equiv \mathbf{A}/(B_{c2}\,\xi)\).
Likewise, the dimensionless heat current density assumes the form
\begin{equation}
\mathbf{j}^{h}
=
-\frac{1}{2}
\left[
\left(\bar{\mathcal{D}}_{\tau}\psi\right)^{*}\,\bar{\mathcal{D}}\psi
+
\left(\bar{\mathcal{D}}_{\tau}\psi\right)\,\left(\bar{\mathcal{D}}\psi\right)^{*}
\right].
\end{equation}

Finally, the electrical transport properties are measured in units of the
Ginzburg--Landau conductivity,
\begin{equation}
\sigma_{GL}=\frac{\gamma\,\xi^{2}}{\mu_{0}\,\lambda^{2}}.
\end{equation}



\subsection{Linearization and self-consistent Gaussian approximation}The inherent nonlinearity of the TDGLE poses a significant challenge for analytical treatment, particularly when driven by Langevin thermal noise \cite{schmid1966time}. Nevertheless, for investigations focused on quadratic thermal observables—specifically the superfluid density and electric current—the self-consistent fluctuation approximation provides a reliable and computationally efficient alternative \cite{tinh2009theory, tinh2010electrical, jiang2014strong}.This framework, which draws a close parallel to the Hartree-Fock method in fermionic theory, was extensively formalized in the foundational studies of Rosenstein and collaborators \cite{kovner1989covariant, rosenstein1989covariant, wang2017covariant}. The approach involves linearizing the cubic interaction term $|\psi|^2 \psi$ through the substitution $2\langle |\psi|^2 \rangle \psi$, a procedure that effectively captures the essential many-body correlations while maintaining mathematical tractability. Within the self-consistent Hartree scheme, the stochastic dynamics of the
order parameter is governed by a linearized dimensionless TDGL equation.
For stationary and spatially homogeneous processes, the superfluid density
reduces to a constant, and the equation of motion can be written as
\begin{equation}
\left(
\overline{\mathcal{D}}_{\tau}
-\frac{1}{2}\,\overline{\boldsymbol{\mathcal{D}}}^{\,2}
-\frac{\mathcal{B}}{2}
-\frac{v^{2}}{2}
+\epsilon
\right)\psi(\mathbf{r},\tau)
=
\overline{\zeta}(\mathbf{r},\tau),
\label{eq:tdgle_final}
\end{equation}
where $v=\mathcal{E}/\mathcal{B}$ denotes the drift velocity induced by the
electric field. The self-consistently determined fluctuation gap $\epsilon$
collects all energy shifts due to temperature, magnetic field, and electric
field, and is given by
\begin{equation}
\epsilon
=
\frac{\mathcal{T}-1+\mathcal{B}+v^{2}}{2}
+
2\left\langle |\psi|^{2} \right\rangle .
\label{eq:epsilon_final}
\end{equation}
This shifted formulation guarantees that the fluctuation gap remains positive
even in the presence of a finite electric field, and provides a convenient
starting point for the Green's-function analysis.

The formal solution is the convolution
\begin{equation}
\psi(\mathbf{r}_1,\tau_1)
=
\int d\mathbf{r}_2 \int d\tau_2\;
G(\mathbf{r}_1,\tau_1;\mathbf{r}_2,\tau_2)\,
\overline{\zeta}(\mathbf{r}_2,\tau_2),
\label{eq:psi_conv}
\end{equation}

Using the covariant Gaussian construction introduced above, the Green's function
$G(\mathbf r_1,\tau_1;\mathbf r_2,\tau_2)$ associated with the linearized TDGL operator
can be obtained in closed form. Specifically, $G$ satisfies the inhomogeneous equation
\begin{equation}
\Bigg(
  \overline{\mathcal{D}}_{\tau_1}
  - \frac{1}{2}\,\overline{\boldsymbol{\mathcal{D}}}_{1}^{\,2}
  - \frac{\mathcal{B}}{2}
  - \frac{v^{2}}{2}
  + \epsilon
\Bigg)
G(\mathbf r_1,\tau_1;\mathbf r_2,\tau_2)
\nonumber=
2\omega t\,
\delta(\mathbf r_1-\mathbf r_2)\,
\delta(\tau_1-\tau_2),
\label{eq:green_eq_shift_1}
\end{equation}
where the operators $\overline{\mathcal{D}}_{\tau_1}$ and 
$\overline{\boldsymbol{\mathcal{D}}}_{1}$ act on the coordinates $(\mathbf r_1,\tau_1)$ of the Green's function.

For a positive time separation $\tau=\tau_1-\tau_2>0$, the Green function admits a compact
factorized form,
\begin{equation}
G\!\left(\mathbf r_1,\tau_1;\mathbf r_2,\tau_2\right)
=
C(\tau)\,
\exp\!\left[
\mathrm i\,\frac{\mathcal B}{2}(y_1+y_2)\,X
\right]
\exp\!\left(
-\frac{X^{2}+Y^{2}}{2W(\tau)}-vX
\right),
\label{eq:G_factorized}
\end{equation}
where the relative variables are defined as
\begin{equation}
\tau=\tau_1-\tau_2,\qquad
X=x_1-x_2-v\tau,\qquad
Y=y_1-y_2 .
\label{eq:XYtau_def}
\end{equation}
The prefactor $C(\tau)$ encodes the full temporal decay and Landau-level structure
of the fluctuating Cooper pairs and is given explicitly by
\begin{equation}
C(\tau)
=
\frac{\mathcal B}{4\pi}\,
\frac{
\exp\!\left[-\left(\epsilon-\dfrac{\mathcal B}{2}\right)\tau\right]
}{
\sinh\!\left(\dfrac{\mathcal B}{2}\,\tau\right)
}.
\label{eq:Ctau_def}
\end{equation}

The Gaussian width $W(\tau)$ characterizes the spatial spreading of the fluctuation
packet under the combined action of magnetic confinement and electric-field drift.
It is uniquely determined by the Gaussian ansatz together with the regularity
condition $W(0)=0$, yielding
\begin{equation}
W(\tau)
=
\frac{2}{\mathcal B}\,
\tanh\!\left(\frac{\mathcal B}{2}\,\tau\right).
\label{eq:W_definition}
\end{equation}
The overall normalization of $G$ is fixed by the $\delta$-function source term in
the defining Green-function equation.

\section{Fluctuation Cooper-pair density and transport properties}

\subsection{Fluctuation Cooper-pair density}

The presence of superconducting fluctuations is quantified by the
noise-averaged amplitude of the order parameter,
$\langle |\psi|^{2} \rangle$, which measures the density of
fluctuation-induced Cooper pairs.
Making use of the Green's function in
Eq.~\eqref{eq:G_factorized}, together with the white-noise correlator, this
quantity can be expressed as
\begin{equation}
\bigl\langle |\psi(\mathbf{r},\tau)|^{2} \bigr\rangle
=
\frac{\omega\,\mathcal{B}\,\mathcal{T}}{2\pi}
\int_{\tau_c}^{+\infty}
\frac{f(\epsilon,\tau)}
{\sinh\!\left(\mathcal{B}\tau\right)}
\,\mathrm{d}\tau .
\label{eq:psi2_avg}
\end{equation}

The dimensionless kernel $f(\epsilon,\tau)$ encapsulates the combined
influence of the fluctuation gap, magnetic confinement, and the drift
induced by the applied electric field. Its explicit form reads
\begin{equation}
f(\epsilon,\tau)
=
\exp\!\left[
\left(-2\epsilon+\mathcal{B}\right)\tau
\right]
\exp\!\left[
\frac{2v^{2}}{\mathcal{B}}
\tanh\!\left(\frac{\mathcal{B}\tau}{2}\right)
\right],
\label{eq:f_kernel}
\end{equation}
where $v=\mathcal{E}/\mathcal{B}$ denotes the electric-field-induced drift
velocity.

The fluctuation gap $\epsilon$, which controls the strength of
superconducting fluctuations, is determined self-consistently.
Combining Eqs.~\eqref{eq:epsilon_final} and \eqref{eq:psi2_avg}, one
obtains the integral equation
\begin{equation}
\epsilon
=
\frac{\mathcal{T}-1+\mathcal{B}+v^{2}}{2}
+
\frac{\omega\,\mathcal{B}\,\mathcal{T}}{\pi}
\int_{\tau_c}^{+\infty}
\frac{f(\epsilon,\tau)}
{\sinh\!\left(\mathcal{B}\tau\right)}
\,\mathrm{d}\tau ,
\label{eq:epsilon_sc}
\end{equation}
where $\tau_c$ represents the ultraviolet cutoff arising from the
short-time regularization of the TDGL theory.

\subsection{Electric current and renormalized resistance}

When an external electric field is applied, the fluctuating Cooper pairs
are driven out of equilibrium and give rise to a finite superconducting
contribution to transport.
After averaging over the Langevin noise, the fluctuation-induced
supercurrent density along the field direction ($y$ axis) is given by
\begin{equation}
\bigl\langle j^{s}_{y} \bigr\rangle
=
\frac{\omega\,\mathcal{E}\,\mathcal{T}}{4\pi}
\int_{0}^{+\infty}
\frac{f(\epsilon,\tau)}
{\cosh^{2}\!\left(\dfrac{\mathcal{B}\tau}{2}\right)}
\,\mathrm{d}\tau .
\label{eq:js_avg}
\end{equation}

In a two-dimensional superconducting thin film, the total longitudinal
current density results from the coexistence of a normal dissipative
channel, $J_n=\sigma_n E$, and a fluctuation-induced superconducting
contribution.
Accordingly, the noise-averaged current density along the $y$ direction
can be written as
\begin{equation}
\bigl\langle J_y \bigr\rangle
=
\sigma_n E
\left[
1
+
\frac{\omega\,\mathcal{T}}{4\pi k}
\int_{0}^{+\infty}
\frac{f(\epsilon,\tau)}
{\cosh^{2}\!\left(\dfrac{\mathcal{B}\tau}{2}\right)}
\,\mathrm{d}\tau
\right],
\label{eq:J_total}
\end{equation}
where $k=\sigma_n/\sigma_{\mathrm{GL}}$ denotes the ratio of the
normal-state conductivity to the Ginzburg--Landau conductivity.

Introducing the normal-state resistance $R_n$, the
renormalized resistance follows in the compact form
\begin{equation}
R
=
\frac{R_n}{
1
+
\dfrac{\omega\,\mathcal{T}}{4\pi k}
\displaystyle
\int_{0}^{+\infty}
\frac{f(\epsilon,\tau)}
{\cosh^{2}\!\left(\dfrac{\mathcal{B}\tau}{2}\right)}
\,\mathrm{d}\tau
},
\label{eq:R_renorm}
\end{equation}
which makes explicit how interacting superconducting fluctuations
renormalize the electrical transport response.
\subsection{Heat current density and  the
transverse thermoelectric coefficient}

Beyond charge transport, the combined action of electric-field-induced drift and magnetic confinement generates a transverse energy flow. 
After averaging over the Langevin noise, the $x$ component of the dimensionless heat-current density reads
\begin{equation}
\bigl\langle j^{h}_{x} \bigr\rangle
=
\frac{\omega\,\mathcal{T}\,\mathcal{E}\,\mathcal{B}}{4\pi}
\int_{0}^{+\infty}
\frac{f(\epsilon,\tau)}
{\cosh^{2}\!\left(\dfrac{\mathcal{B}\tau}{2}\right)}
\,\mathrm{d}\tau ,
\label{eq:jhx_avg}
\end{equation}
where the kernel $f(\epsilon,\tau)$ is defined in Eq.~\eqref{eq:f_kernel}. 
Equation~\eqref{eq:jhx_avg} therefore represents the fluctuation-induced transverse heat current in rescaled (dimensionless) units, normalized by the characteristic Ginzburg--Landau heat-current scale.

To restore physical units, the dimensionless quantity in
Eq.~\eqref{eq:jhx_avg} must be multiplied by the heat-current scale
$J_{GL}^{h}$ defined in Eq.~\eqref{eq:JGL_h}.
Using the linear-response relation between the transverse heat current
and the applied electric field (see
Eq.~\eqref{eq:alpha_xy_definition}), the transverse thermoelectric
(Ettingshausen) coefficient is obtained as
\begin{equation}
\alpha_{xy}
=
\frac{\omega\, B}{8\pi\,\mu_{0}\,\kappa^{2}\,T_{c0}}
\int_{0}^{+\infty}
\frac{f(\epsilon,\tau)}
{\cosh^{2}\!\left(\dfrac{\mathcal{B}\tau}{2}\right)}
\,\mathrm{d}\tau .
\label{eq:alpha_xy_final}
\end{equation}

\section{Numerical results and comparison with experiment}
\subsection{Fluctuation-controlled resistive transitions under out-of-plane magnetic fields}

The temperature dependence of the electrical resistance in the presence of a magnetic field applied perpendicular to the film plane constitutes a stringent and direct test of fluctuation-driven superconductivity in reduced dimensions. Within the time-dependent Ginzburg--Landau (TDGL) framework, thermal Cooper-pair fluctuations treated self-consistently at the Gaussian (Hartree) level lead to a renormalization of the effective Ginzburg--Landau mass. This renormalization produces an intrinsically smooth resistive crossover extending over a broad temperature range, without requiring any extrinsic inhomogeneity or additional transport channels. As demonstrated below, this fluctuation-based mechanism provides a unified description of a wide variety of experimental $R(T)$ data across distinct material platforms [see \cite{lu2015evidence,zhang2023spin,ienaga2024broadened,shao2025anomalous,wahlberg2026boosting}).

A paradigmatic example of fluctuation-dominated transport is provided by gated monolayer MoS$_2$, as reported by Lu \textit{et al.}~\cite{lu2015evidence}. 
In this strictly two-dimensional superconducting system, the resistive transition under an out-of-plane magnetic field shifts continuously toward lower temperatures and becomes markedly broadened, with the resistance decreasing smoothly rather than collapsing abruptly. 
This behavior indicates the persistence of superconducting correlations well above the nominal critical temperature and reflects the dominant role of superconducting fluctuations in low dimensions. 
Within the self-consistent Gaussian TDGL framework, the full set of experimental $R(T,B)$ curves can be quantitatively described using a single parameter set: a critical temperature $T_{c0}\simeq7.5\,\mathrm{K}$, an upper critical field ${B}_{c2}\simeq12\,\mathrm{T}$, a conductivity ratio $k\simeq0.15$, a normal-state resistance $R_n\simeq1233.47\,\Omega$, and a fluctuation strength $\omega\simeq8.5\times10^{-3}$. 
The relatively large value of $\omega$ highlights the enhanced importance of thermal fluctuations in this atomically thin material, where short-lived Cooper pairs generate a substantial excess conductivity over an extended temperature range, fully consistent with the fluctuation-controlled superconductivity scenario established in Ref.~\cite{lu2015evidence}.

\FloatBarrier
\begin{figure}[H]
  \centering
\includegraphics[width=0.8\linewidth]
  {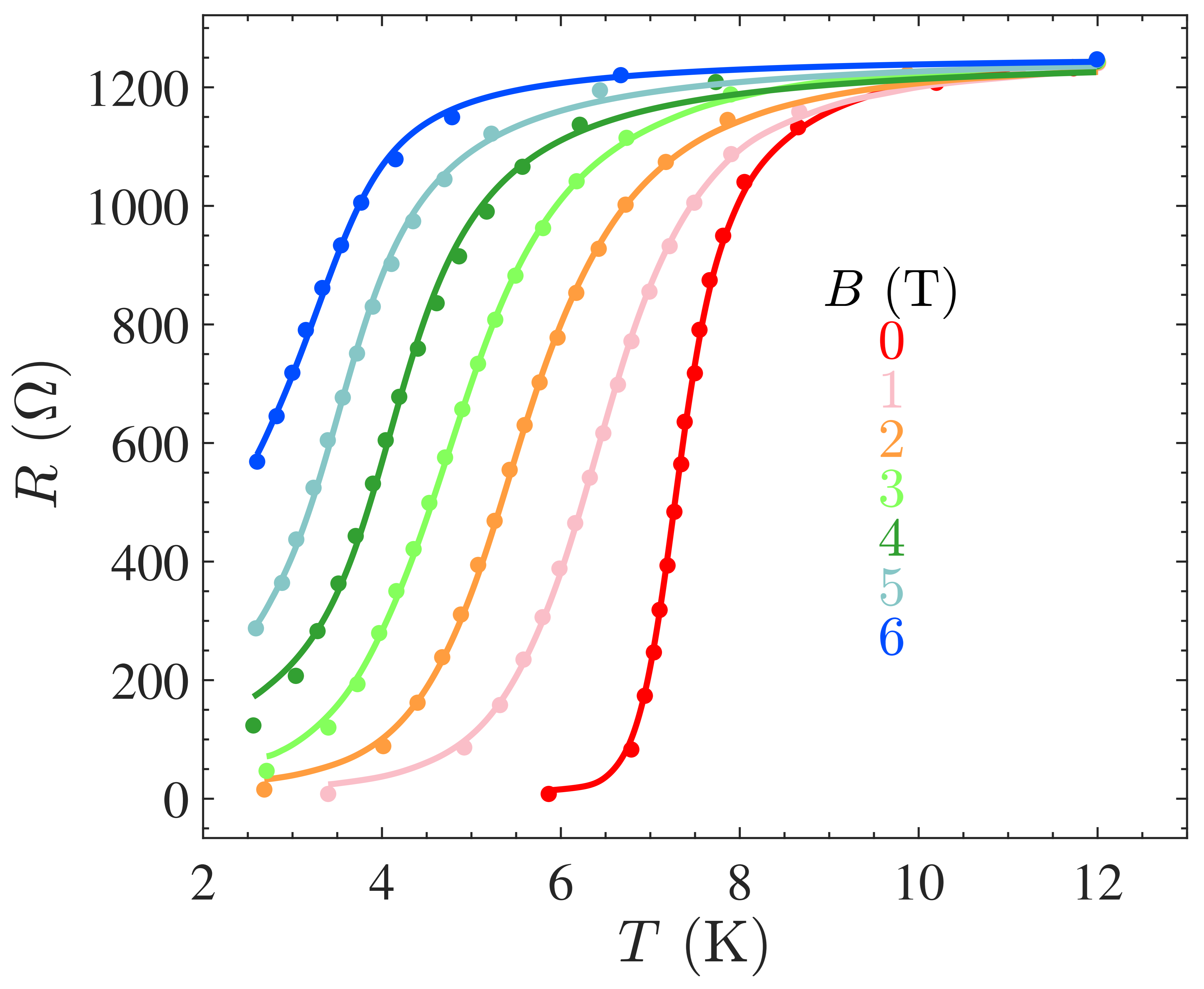}
  \caption{Electrical resistance $R(T)$ of gated monolayer MoS$_2$ under out-of-plane magnetic fields, reproduced from Ref.~\cite{lu2015evidence}. The field-driven shift and systematic broadening of the transition reflect strong superconducting fluctuations in two dimensions.}
  \label{fig:RT_MoS2_2015}
\end{figure}
\FloatBarrier
A closely related fluctuation-controlled resistive phenomenology is observed in atomically thin 2M-WS$_2$, as reported by Zhang \textit{et al.}~\cite{zhang2023spin}. 
Although this system hosts strong spin--orbit--parity coupling, the application of a perpendicular magnetic field produces a smooth and continuous evolution of the resistive crossover rather than a sharp transition. 
Within the self-consistent Gaussian TDGL framework, the full set of experimental $R(T,B)$ data is quantitatively described using a single parameter set: a critical temperature $T_{c0}\simeq7.29\,\mathrm{K}$, an upper critical field ${B}_{c2}\simeq8.46\,\mathrm{T}$, a conductivity ratio $k\simeq0.0195$, and a fluctuation strength $\omega\simeq1.26\times10^{-3}$. 
The close correspondence with gated monolayer MoS$_2$ demonstrates that the pronounced broadening of the resistive transition is largely insensitive to microscopic pairing details and spin--orbit structure, and is instead governed by the universal dynamics of interacting superconducting fluctuations captured by the self-consistent TDGL theory.

\FloatBarrier
\begin{figure}[H]
  \centering
\includegraphics[width=0.9\linewidth]
  {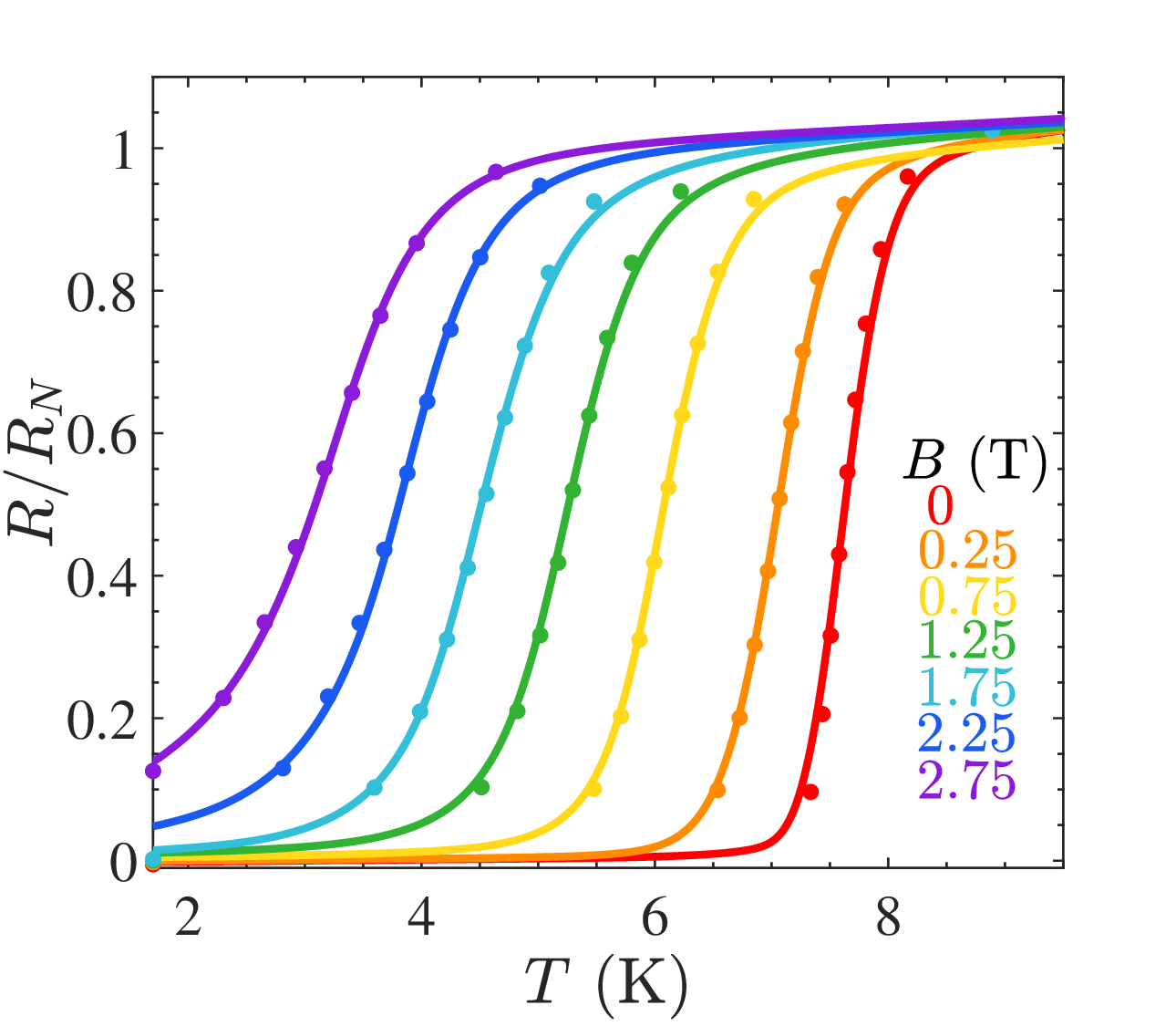}
  \caption{Resistance $R(T)$ of atomically thin 2M-WS$_2$ under out-of-plane magnetic fields, reproduced from Ref.~\cite{zhang2023spin}. The continuous crossover highlights the robustness of fluctuation-dominated transport in low-dimensional systems with strong spin--orbit coupling.}
  \label{fig:RT_WS2_2023}
\end{figure}
\FloatBarrier

The quantitative predictive power of the self-consistent Gaussian TDGL approximation becomes particularly evident in strongly disordered superconducting thin films, as demonstrated in the recent report by Ienaga \textit{et al.}~\cite{ienaga2024broadened}. 
Using the experimentally determined normal-state resistance $R_n \simeq250.14\,\Omega$, the full set of measured $R(T)$ curves is quantitatively reproduced with a single, internally consistent parameter set: a critical temperature $T_{c0} \simeq 2.47\,\mathrm{K}$, an upper critical field ${B}_{c2} \simeq  6.04\,\mathrm{T}$, a conductivity ratio $k \simeq0.02$, and a fluctuation strength $\omega  \simeq 1.05\times10^{-4}$. 
Within the TDGL--Gaussian framework, the exceptionally broad resistive transition and its smooth evolution under increasing perpendicular magnetic field arise naturally from the strong self-consistent renormalization of the superconducting fluctuation spectrum near a disorder-driven quantum critical regime. 
Importantly, this level of agreement is achieved without invoking additional assumptions beyond interacting thermal fluctuations, in direct correspondence with the fluctuation-controlled scenario established in Ref.~\cite{ienaga2024broadened}.

\FloatBarrier
\begin{figure}[H]
  \centering
  \includegraphics[width=\linewidth,keepaspectratio]{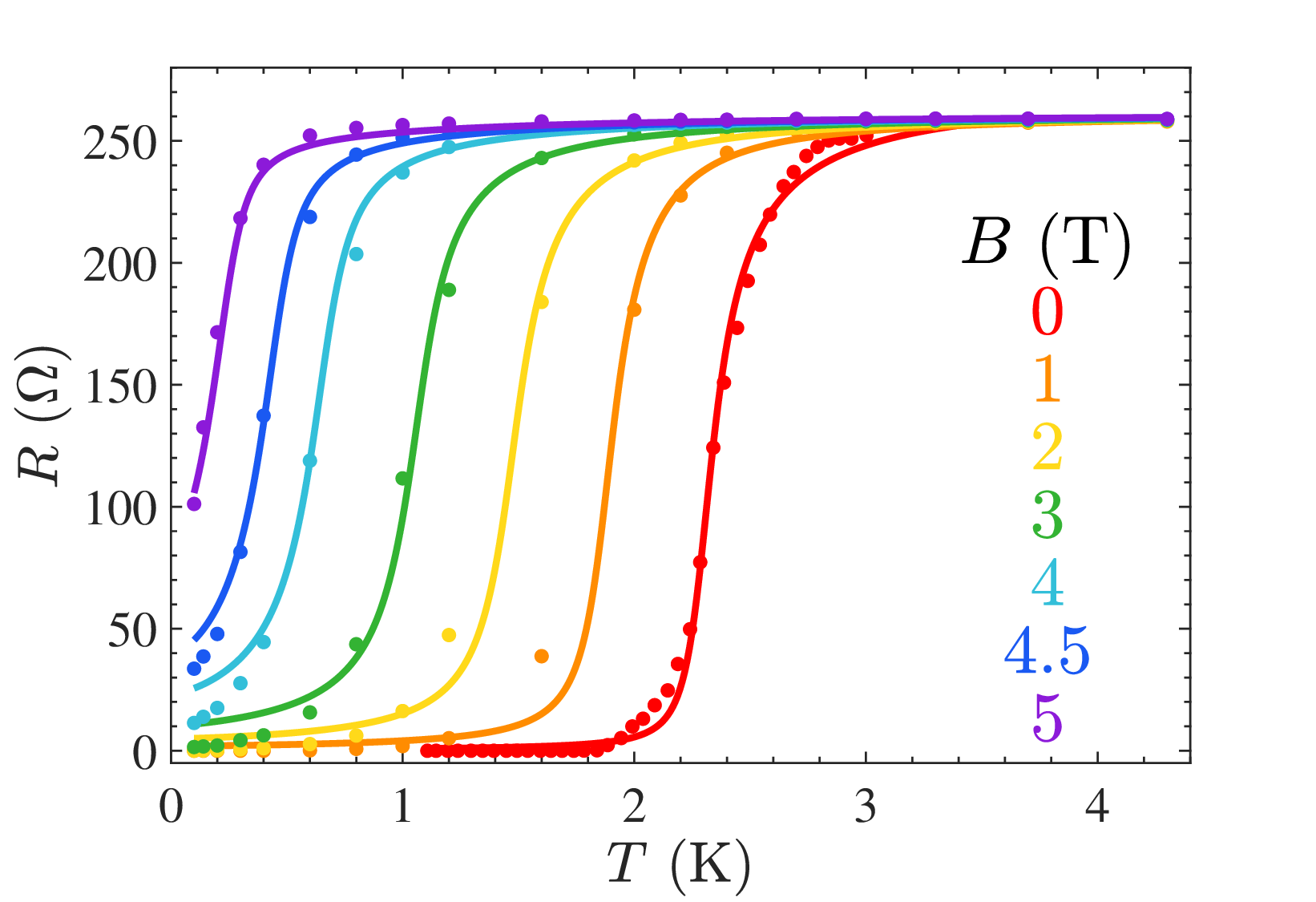}
\caption{Electrical resistance $R(T)$ of a disordered superconducting thin film under out-of-plane magnetic fields, reproduced from Ref.~\cite{ienaga2024broadened}. Symbols denote experimental data, while solid lines represent theoretical fits obtained within the self-consistent Gaussian TDGL approach.}
\label{fig:RT_NC_2024}
\end{figure}
\FloatBarrier

A further manifestation of this universal, fluctuation-controlled transport behavior is found in underdoped infinite-layer nickelate superconducting thin films, as reported in~\cite{shao2025anomalous}. 
Despite their unconventional electronic structure and proximity to an anomalous metallic regime, the resistive transitions under out-of-plane magnetic fields display the same smooth crossover and pronounced broadening characteristic of fluctuation-dominated superconductivity. 
Within the self-consistent Gaussian TDGL framework, the experimental $R(T,B)$ data are consistently captured using a single parameter set: a critical temperature $T_{c0}\simeq5.00\,\mathrm{K}$, an upper critical field ${B}_{c2}\simeq13.7\,\mathrm{T}$, a conductivity ratio $k\simeq0.0578$, a normal-state resistance $R_n\simeq272.7\,\Omega$, and a comparatively large fluctuation strength $\omega\simeq9.07\times10^{-2}$. 
The sizable value of $\omega$ reflects the extreme strength of superconducting fluctuations in the nickelate system, underscoring that fluctuation-induced conductivity dominates the transport response and placing infinite-layer nickelates within the same fluctuation universality class as transition-metal dichalcogenides and strongly disordered superconducting thin films.

\FloatBarrier
\begin{figure}[H]
  \centering
  \includegraphics[width=\linewidth,keepaspectratio]{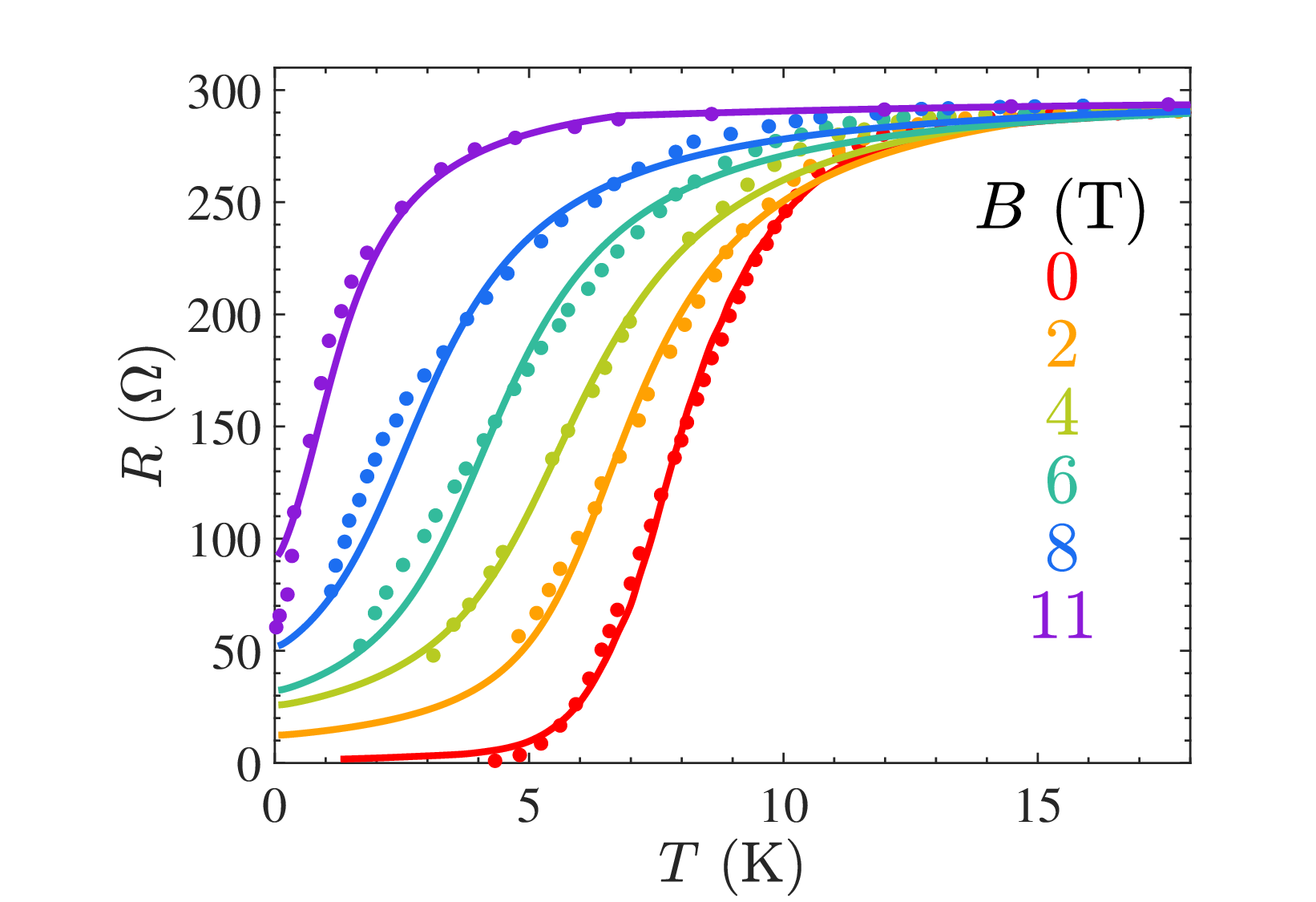}
\caption{Resistance $R(T)$ of underdoped infinite-layer nickelate superconducting thin films under out-of-plane magnetic fields, reproduced from Ref.~\cite{shao2025anomalous}. The broadened transitions emphasize the universality of fluctuation-driven transport beyond cuprate-based systems.}
\label{fig:RT_Nickelate_2025}
\end{figure}
\FloatBarrier
The quantitative agreement obtained across distinct material platforms—including monolayer transition-metal dichalcogenides, strongly disordered thin films, and infinite-layer nickelates—demonstrates that the smooth magnetic-field-driven broadening of the resistive transition is a universal fingerprint of interacting superconducting fluctuations in reduced dimensions. 
In each case, the full manifold of experimental $R(T,B)$ data is consistently reproduced using a single, internally coherent parameter set within the self-consistent Gaussian TDGL framework. 
This robustness indicates that the observed resistive crossover does not originate from extrinsic inhomogeneity or material-specific microscopic details, but rather from fluctuation-renormalized Cooper-pair dynamics intrinsic to low-dimensional superconductors. 
The fluctuation strength parameter $\omega$ quantitatively captures the degree of thermal and quantum fluctuation effects, providing a unified scale for comparing diverse superconducting systems. 
Taken together, these results establish fluctuation-controlled superconductivity as a universal emergent transport regime in low-dimensional correlated materials.

\subsection{S-shaped $I$--$V$ characteristics as a historical and theoretical hallmark of nonequilibrium superconductivity}

S-shaped current--voltage ($I$--$V$) or current--electric-field ($J$--$E$) characteristics constitute one of the most distinctive nonlinear fingerprints of driven superconductivity. 
An S-shaped curve is defined by the existence of a negative-differential segment ($dI/dV<0$ or $dJ/dE<0$), which implies multivalued transport solutions and an intrinsic dynamical instability under current bias: the unstable branch is generically bypassed by voltage switching, whereas a genuine voltage-controlled protocol is required to stabilize and resolve the full nonlinear response. 
This basic stability constraint explains why, despite the century-long history of superconductivity since Onnes' discovery of vanishing resistance in 1911~\cite{onnes1911resistance}, a ``clean'' experimental observation of a continuous S-shaped branch has remained rare and has repeatedly triggered debates on its microscopic origin.

The modern nonlinear-transport history of S-shapes begins with the seminal by Vodolazov et al.~\cite{vodolazov2003current}, who demonstrated within the time-dependent Ginzburg--Landau (TDGL) framework~\cite{schmid1966time} that quasi-one-dimensional superconductors admit multivalued transport solutions in the constant-voltage regime, yielding an ``unusual S behavior'' as an intrinsic superconducting state rather than a measurement artifact. 
Soon after, nanowire experiments directly visualized how localized suppression of the order parameter evolves into phase-slip centers and abrupt voltage jumps~\cite{michotte2004development}, thereby clarifying why current-biased measurements often show switching instead of a resolvable S-branch. 
The broader relevance of nonlinear switching and S-shaped tendencies was then recognized in correlated and higher-dimensional superconductors: nanoscale transport in cuprates indicated that electrical drive can destroy phase coherence while leaving the pairing gap comparatively intact~\cite{wei2005nanoscale,kaminski2016destroying}, highlighting that strong fluctuations and nonequilibrium superconducting dynamics, rather than trivial Joule heating, can control nonlinear transport.

A decisive conceptual advance was achieved by Qiao \textit{et al.}~\cite{qiao2018dynamical}, who introduced a fluctuation-driven mechanism for S-shaped transport within TDGL theory supplemented by a self-consistent Gaussian (Hartree) treatment of thermal Cooper-pair fluctuations, grounded in the broader fluctuation literature~\cite{schmid1969diamagnetic,larkin2005theory,ullah1991effect,rosenstein2010ginzburg,kovner1989covariant,rosenstein1989covariant}. 
In this framework, the electric field renormalizes the effective Ginzburg--Landau mass via feedback from fluctuating Cooper pairs, so that the superconducting contribution to the current becomes intrinsically nonmonotonic in $E$, producing $dJ/dE<0$ even at $B=0$ and without invoking vortices. 
The instability boundary is predicted quantitatively by the simultaneous conditions $dJ/dE=0$ and $d^{2}J/dE^{2}=0$~\cite{qiao2018dynamical}. 
Independent extensions have reinforced and generalized this picture: Dang and Tinh~\cite{dang2022voltage} showed that S-shaped voltage--current characteristics persist under magnetic field and that $B$ suppresses the instability continuously rather than generating it, while a subsequent analysis reported in Physica C~\cite{vi2024fluctuation} revisited the original Vodolazov scenario and demonstrated explicitly that self-consistent fluctuation renormalization alone reproduces S-shaped $I$--$V$ curves in nanowires under electric field. 
Taken together, these works establish a coherent theoretical lineage in which S-shaped transport emerges as an intrinsic consequence of nonequilibrium superconducting fluctuations.

\begin{figure}[H]
  \centering
  \includegraphics[width=\linewidth,keepaspectratio]{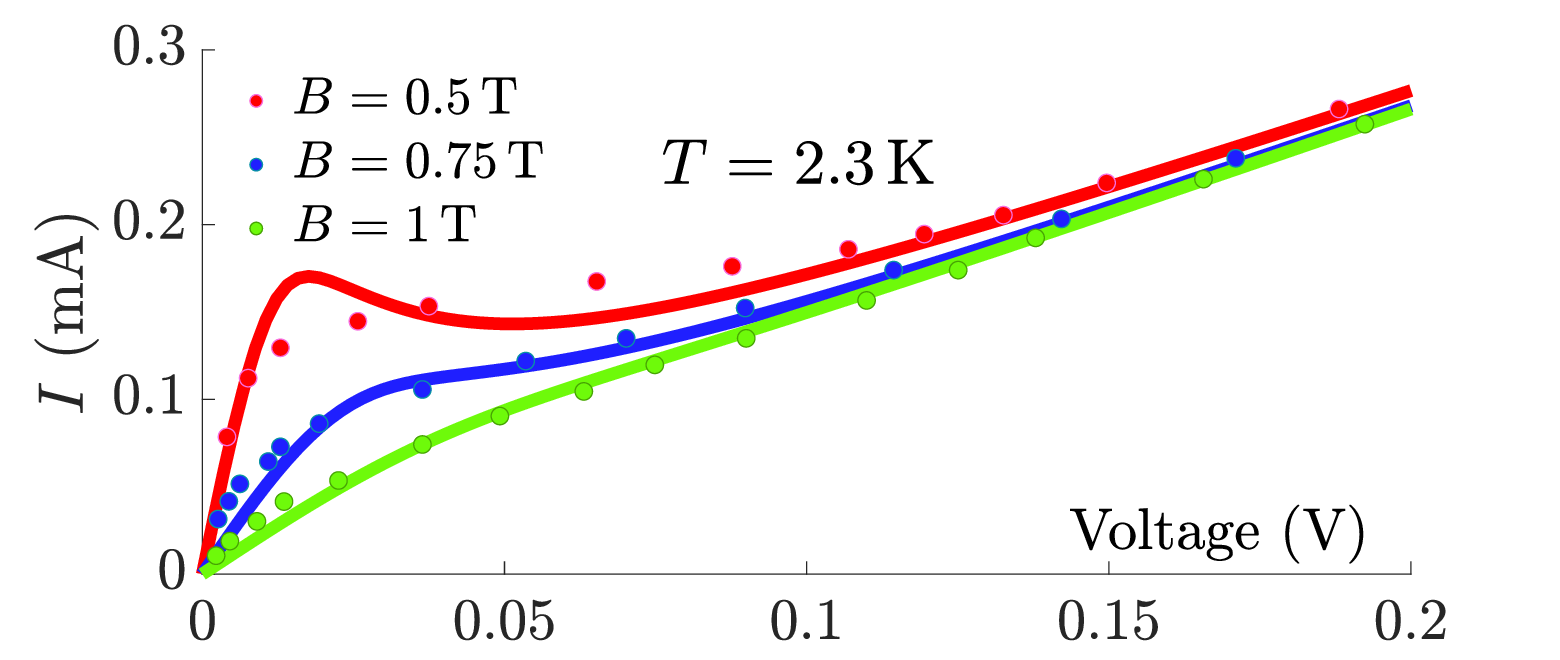}
\caption{Voltage-controlled nonlinear $I$--$V$ characteristics of disordered TiN thin films at various magnetic fields (symbols), together with TDGL--Gaussian fluctuation fits (solid lines), adapted from Ref.~\cite{yadav2021robust}. The emergence of an S-shaped response already at weak field highlights the intrinsic nature of the instability and illustrates why resolving the continuous S-branch requires true voltage control.}
\label{fig:IV_TiN_robust}
\vspace{-2mm}
\end{figure}

Within this historical context, our present numerical analysis provides a stringent experimental test of the fluctuation-driven scenario in a realistic disordered two-dimensional film. 
Figure~\ref{fig:IV_TiN_robust} compares voltage-controlled nonlinear $I$--$V$ characteristics of robustly nitrided TiN thin films~\cite{yadav2021robust} with our TDGL--Gaussian (Hartree) calculations. 
Despite the intrinsic structural disorder introduced by the substrate-mediated nitridation process and the presence of minority silicide phases, the nonlinear transport response is quantitatively reproduced within a single, physically constrained parameter set: a critical temperature $T_{c0} \simeq 3.03\,\mathrm{K}$, an upper critical field $B_{c2}\simeq6.25\,\mathrm{T}$ extracted from the experimentally determined coherence length, a conductivity ratio $k \simeq0.096$, and a fluctuation strength $\omega \simeq 1.32\times10^{-3}$. 
Within this self-consistent Gaussian TDGL framework, the entire nonlinear voltage range across multiple magnetic fields is captured without retuning parameters. 
Remarkably, a clear S-shaped tendency emerges already in the weak-field regime ($B\!\sim\!0.5~\mathrm{T}$, corresponding to $b\sim0.1$ in reduced units) and persists against moderate parameter variations. 

\begin{figure}[t]
  \centering
  \includegraphics[width=\linewidth]{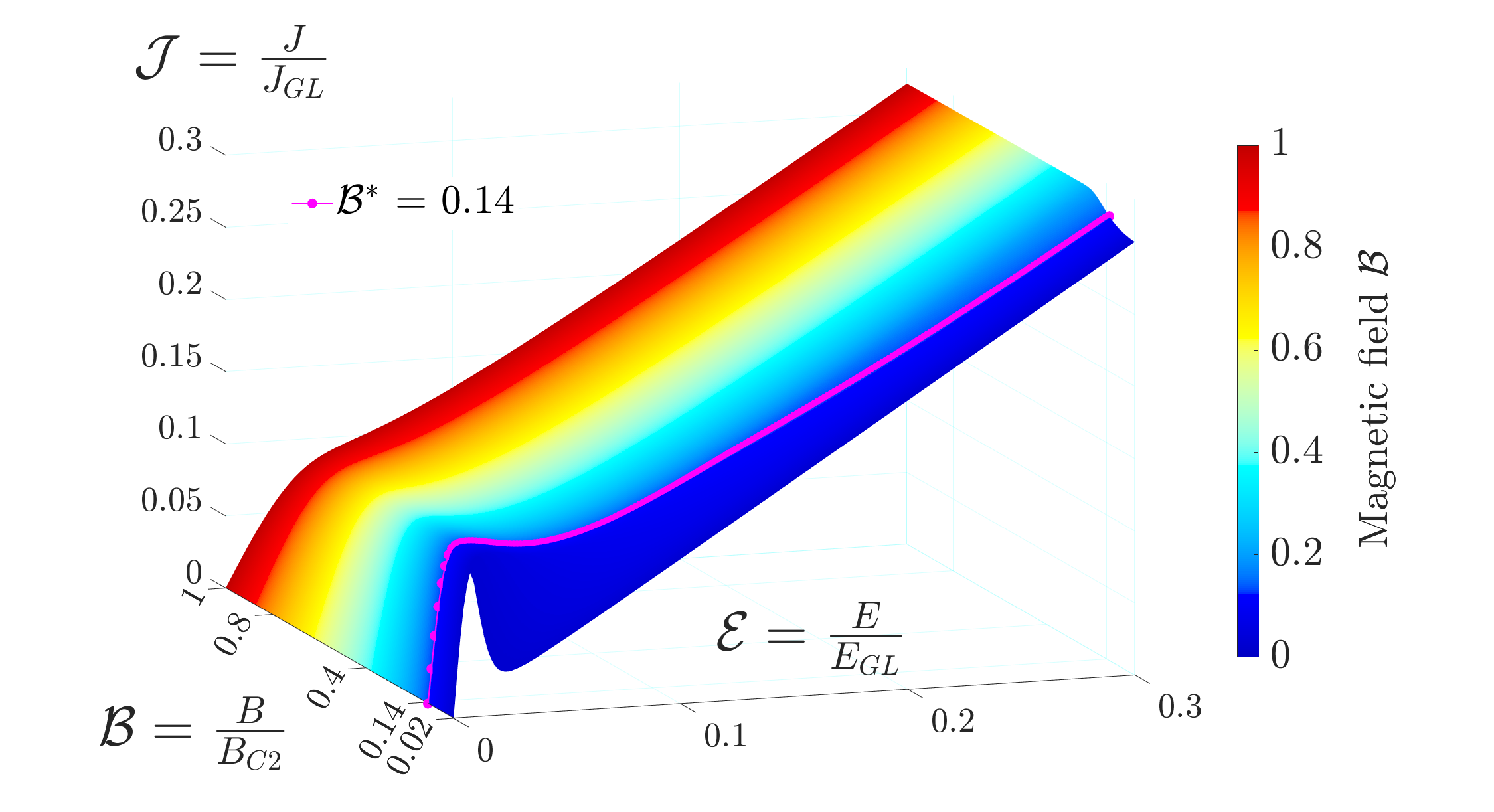}
  \vspace{-2mm}
  \caption{Representative cuts of the calculated $J(E,B)$ manifold at selected characteristic magnetic fields $B^{\ast}$. 
  The folding of the surface corresponds to the S-shaped instability ($dJ/dE<0$). 
  Increasing magnetic field progressively suppresses the fold and eventually eliminates the multivalued branch.}
  \label{fig:JE_surface_B}
  \vspace{-2mm}
\end{figure}

The predictive power of the theory becomes most transparent when the nonlinear response is represented as a geometrical manifold rather than as isolated curves. 
Figure~\ref{fig:JE_surface_B} shows representative cuts of the calculated $J(E,B)$ manifold at selected characteristic magnetic fields $B^{\ast}$. 
In this representation, the S-shaped response acquires a direct geometrical interpretation: it corresponds to a fold of the $J(E,B)$ surface, i.e., the emergence of multivalued current solutions at fixed $(E,B)$. 
With increasing magnetic field, this fold is progressively flattened and ultimately eliminated. 
The disappearance of the S-shape is therefore not an empirical criterion but a sharp dynamical prediction, identified by the simultaneous conditions
\[
\left.\frac{\partial J(T,E,B)}{\partial E}\right|_{B_\chi}=0,
\qquad
\left.\frac{\partial^{2} J(T,E,B)}{\partial E^{2}}\right|_{B_\chi}=0,
\]
which define the terminal point of the fold where the region of negative differential conductivity ceases to exist.

\FloatBarrier

This construction provides a direct operational realization of the TDGL--Hartree theory at finite magnetic field: increasing $B$ suppresses superconducting fluctuations and continuously removes the instability, in agreement with Ref.~\cite{dang2022voltage}. 

\begin{figure}[t]
  \centering
  \includegraphics[width=\linewidth]{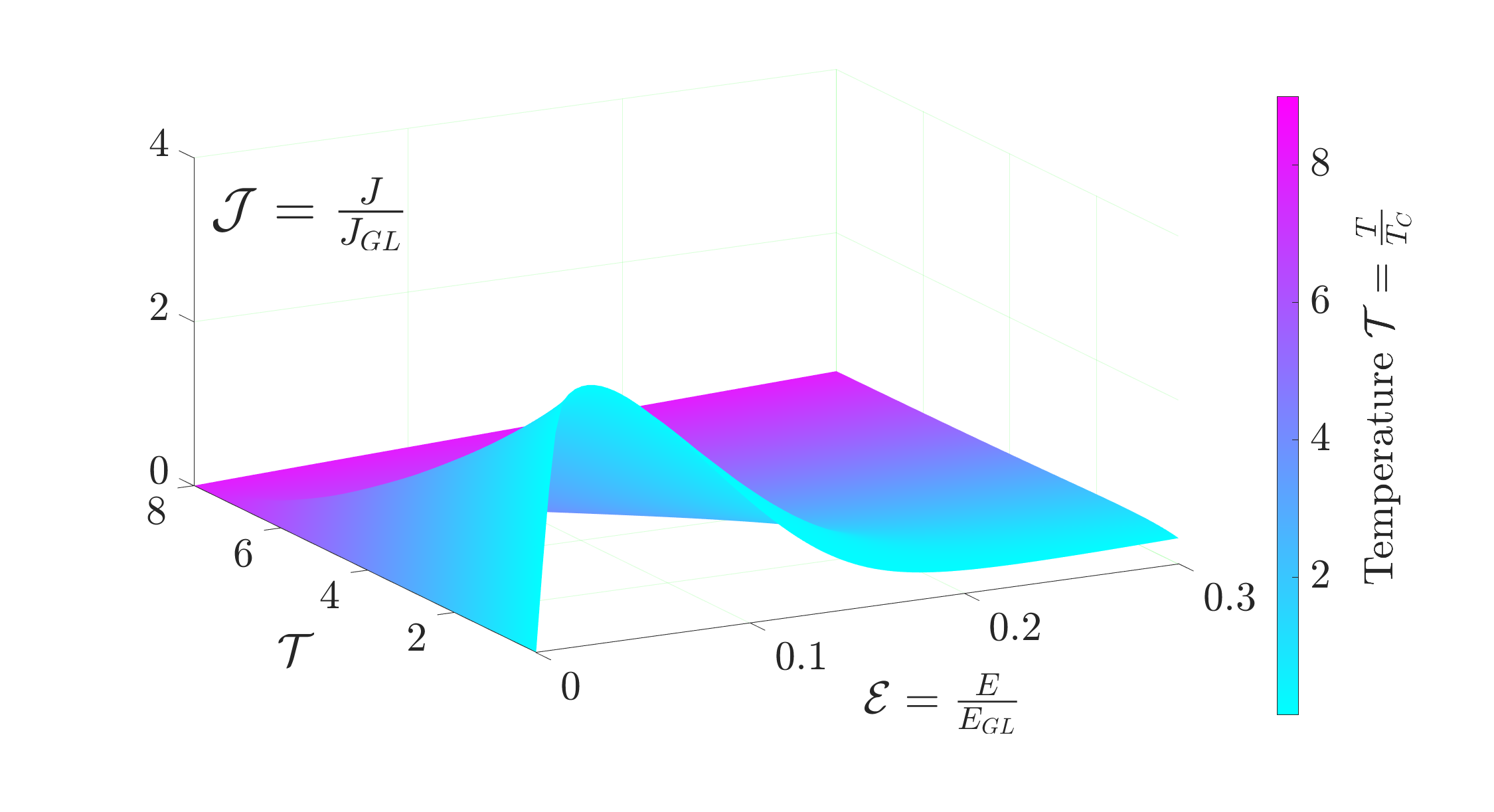}
  \vspace{-2mm}
  \caption{Calculated nonlinear $J(E)$ response over an extended parameter range. 
  The progressive reduction of the instability window corresponds to the smoothing of folds in the underlying $J(E,B)$ manifold.}
  \label{fig:JE_surface_T}
  \vspace{-2mm}
\end{figure}

A complementary view of the same instability mechanism is presented in Fig.~\ref{fig:JE_surface_T}, which displays the calculated nonlinear $J(E)$ response over an extended parameter range. 
The progressive reduction of the nonlinear window and the weakening of the S-shaped tendency are clearly visible as control parameters are tuned. 
In the surface language, this corresponds to the gradual smoothing of folds and ridges that encode multivalued transport solutions.

\FloatBarrier

Together, these representations condense the full nonlinear dataset into a single geometrical framework in which stability boundaries, multivaluedness, and the predicted disappearance of the S-branch are encoded by the topology of the response manifold.

This behavior is pivotal because it indicates that the S-shaped response is not a fine-tuned vortex configuration or a fragile experimental artifact; rather, it is a robust consequence of fluctuation-dominated transport that appears as soon as the stability window of voltage-controlled measurements is opened. 
In particular, the appearance of S-shaped features in the weak-field regime provides a natural explanation for why many ``robust'' datasets may look comparatively featureless under standard current-controlled protocols: even when the underlying instability exists, the measurement typically collapses it into switching, thereby masking the continuous S-branch.

Against this historical and theoretical backdrop, the experimental study conducted in 2025 on voltage-controlled nonlinear transport in a $\mathrm{Bi}_{2}\mathrm{Te}_{3}/\mathrm{Fe(Se,Te)}$ heterostructure~\cite{nagahama2025two} constitutes a decisive step in resolving the intrinsically unstable S-shaped branch of the superconducting $I$--$V$ response. 
By implementing stabilized voltage scans under magnetic field with low-noise four-probe detection and precise field alignment, the experiment directly accessed the continuous multivalued branch that is typically masked by switching in current-biased measurements. 
Within the self-consistent Gaussian TDGL (Hartree) framework, the nonlinear transport response is quantitatively reproduced using a single, physically constrained parameter set obtained as the averaged solution across multiple fitting realizations: a critical temperature $T_{c0} \simeq 14.66\,\mathrm{K}$, an upper critical field $B_{c2}\simeq46.92\,\mathrm{T}$, a conductivity ratio $k \simeq0.1868$, a fluctuation strength $\omega  \simeq 0.02385$, and a normal-state resistance $R_n \simeq 375.58\,\Omega$. 
Without further parameter adjustment, this averaged set captures the full nonlinear voltage dependence and the emergence of a continuous S-shaped branch, placing the experiment on the same theoretical footing as the fluctuation-driven instability scenario discussed in Refs.~\cite{qiao2018dynamical,dang2022voltage,vi2024fluctuation}. 
Although the experimental interpretation emphasizes vortex-flow instability as the proximate phenomenology, the fluctuation-based perspective identifies the S-shaped branch as a stability fingerprint of a multivalued nonlinear superconducting response. 
Within the TDGL--Hartree mechanism, such multivaluedness can arise intrinsically through fluctuation-renormalized dynamics yielding $dJ/dE<0$, even in the absence of externally imposed vortex structures. 
Accordingly, while vortices may modulate the instability in specific geometries, they are not a necessary condition for the S-shaped form itself; rather, the S-shaped behavior is a generic consequence of fluctuation-controlled nonequilibrium superconducting transport.

\begin{figure}[htbp]
  \centering
  \includegraphics[width=\linewidth,keepaspectratio]{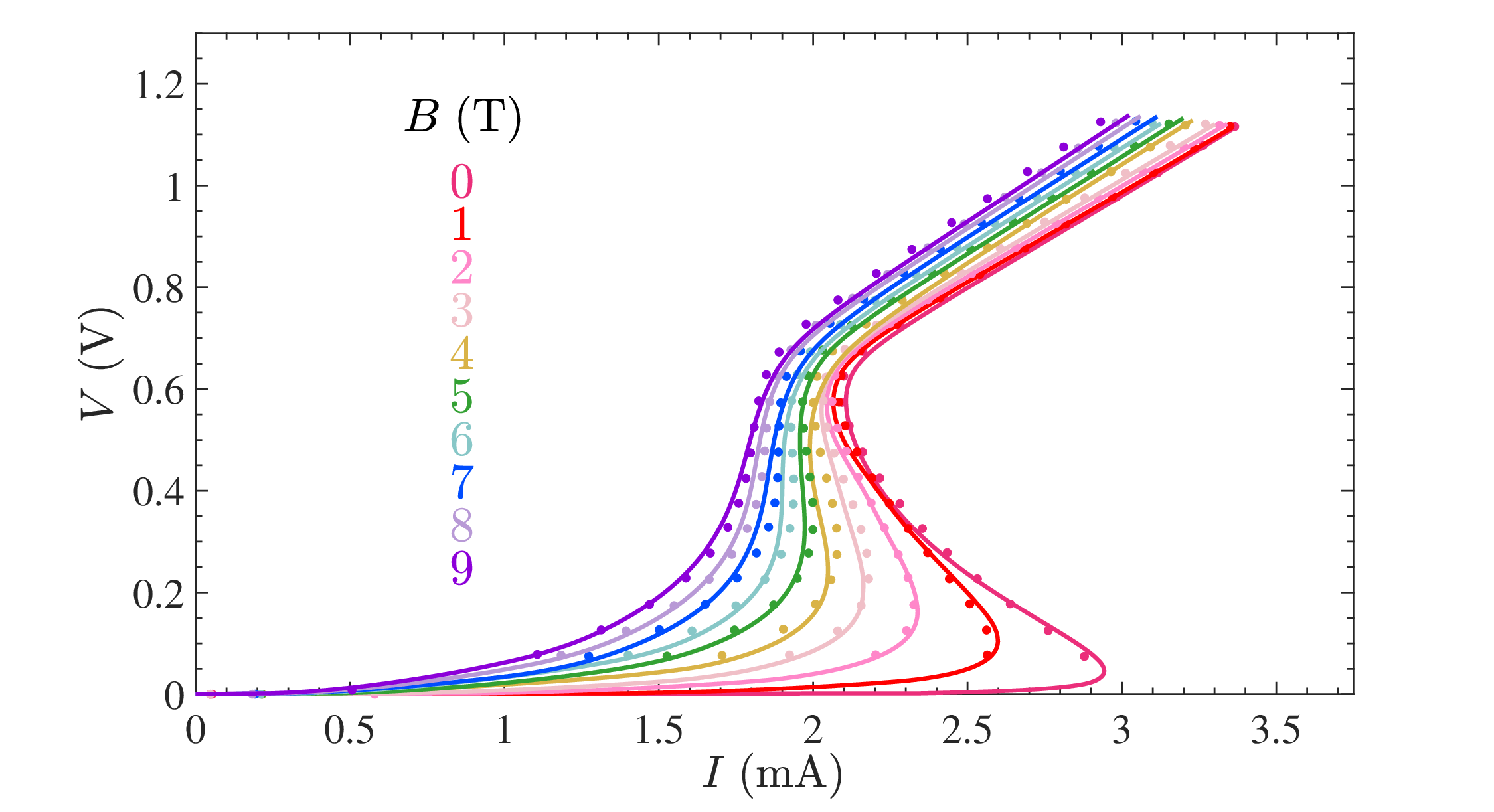}
\vspace{-2mm}
\caption{Comparison between theoretical voltage-scan $V$--$I$ curves and experimental data from the two-dimensional superconducting diode heterostructure reported in Ref.~\cite{nagahama2025two}. The direct resolution of the S-shaped branch under magnetic field represents an exceptional experimental achievement and is quantitatively consistent with a fluctuation-driven multivalued transport response stabilized under voltage control.}
\label{fig:IV_PRL2025_Sshape}
\vspace{-2mm}
\end{figure}

In summary, the evolution from pioneering studies on constant-voltage nanowires~\cite{vodolazov2003current} to recent fluctuation-based nonlinear transport theories and their finite-field extensions~\cite{qiao2018dynamical,dang2022voltage,vi2024fluctuation}---culminating in the recent resolution of S-shaped $I$--$V$ characteristics in 2D heterostructures~\cite{nagahama2025two}---establishes the S-shaped response as a universal hallmark of driven superconductivity across different dimensionalities. 

Our analysis of TiN films and the corresponding surface maps (Figs.~\ref{fig:IV_TiN_robust}--\ref{fig:JE_surface_T}) reinforces this framework. We demonstrate that a fluctuation-renormalized TDGL formalism not only captures diverse experimental datasets with high fidelity but also predicts the emergence and collapse of the S-shaped instability within the control-parameter space. 

Far from being a transient anomaly, the S-shaped response reflects a fundamental interplay between electric-field driving, fluctuation-renormalized correlations, and dynamical stability---a physics that has only become fully accessible through the precision voltage control and noise engineering exemplified in the latest 2025 experiments.
\subsection{Transverse thermoelectric response and vortex entropy: from resistive broadening to $\alpha_{xy}$}

While resistive transitions under out-of-plane magnetic fields provide a sensitive probe of fluctuation-induced conductivity, they do not directly encode the thermodynamic content of the fluctuating superconducting state. In particular, longitudinal transport alone cannot discriminate whether the excess conductivity originates from short-lived amplitude fluctuations, phase-disordered Cooper pairs, or mobile vortex excitations carrying finite entropy. A complementary and conceptually independent probe is therefore required.

A direct and experimentally robust relation between the Nernst signal and superconducting fluctuations is obtained by expressing the transverse thermoelectric response as
\begin{equation}
N = R_{\square}\,\alpha_{xy},
\end{equation}
where $R_{\square}$ denotes the sheet resistance of the two-dimensional film \cite{ienaga2024broadened}. 
This identity is particularly advantageous in strongly disordered superconducting films, where the Nernst signal generated by superconducting fluctuations is known to be orders of magnitude larger than the normal-state contribution. 
As a result, $\alpha_{xy}$ can be reliably extracted from the measured $N$ using the independently determined $R_{\square}$, without contamination from longitudinal thermoelectric effects.

The physical significance of $\alpha_{xy}$ becomes particularly transparent when viewed through its Onsager-reciprocal counterpart, the Ettingshausen effect. In this reciprocal process, a transverse heat current $J_x^{h}$ is generated by an applied electric field $E_y$ in a magnetic field,
\[
J_x^{h}=\alpha_{xy} T E_y.
\]
Thus, $\alpha_{xy}$ directly measures the entropy transported per unit charge in the transverse direction. In the superconducting fluctuation regime, this entropy transport is naturally associated with vortex-like excitations: each mobile vortex carries a finite entropy, and its drift under external fields gives rise simultaneously to both the Nernst and Ettingshausen responses. Consequently, $\alpha_{xy}$ provides a thermodynamic fingerprint of vortex entropy and fluctuation dynamics, rather than merely an additional transport coefficient.

\begin{figure}[H]
  \centering
  \includegraphics[width=\linewidth,keepaspectratio]{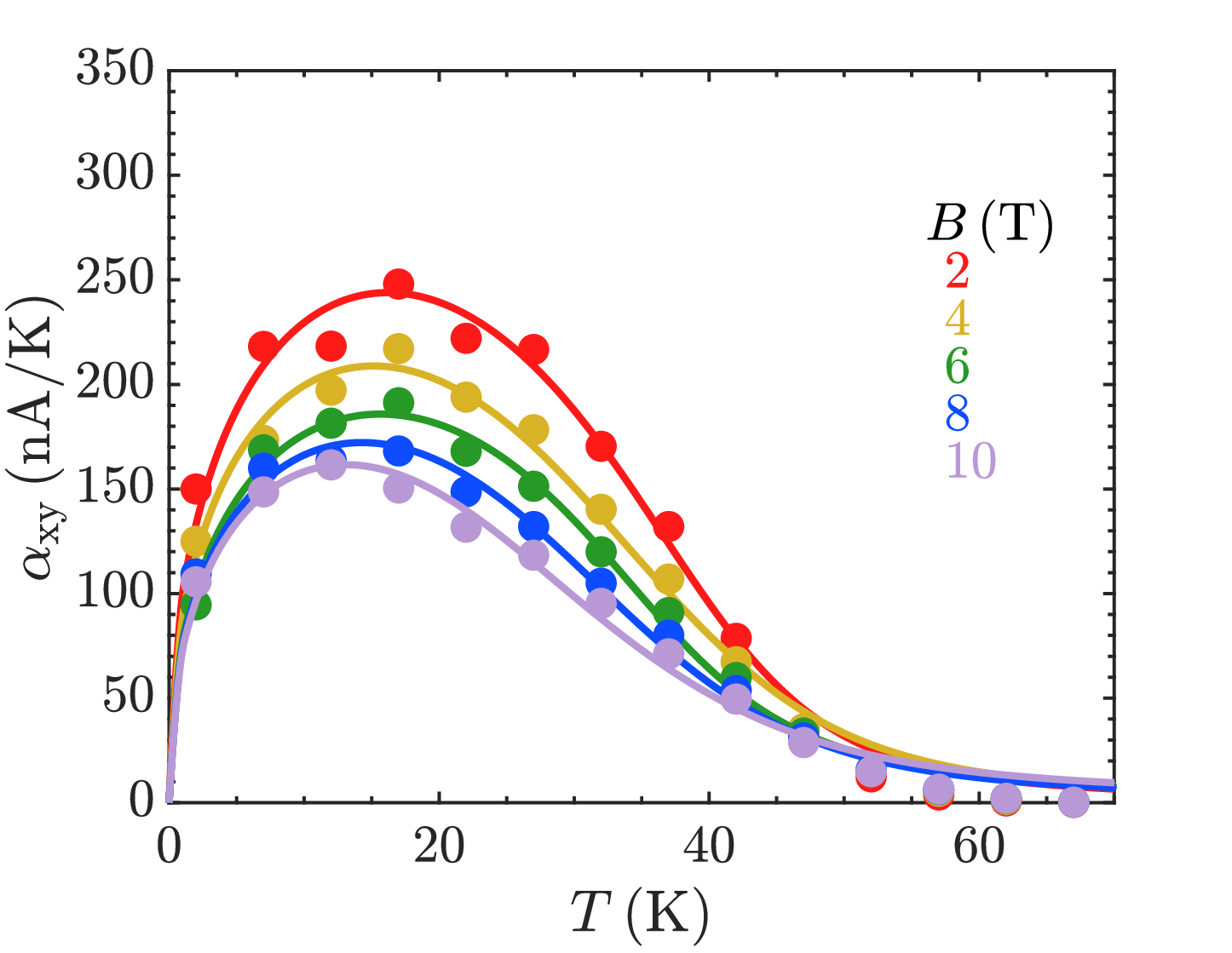}
  \vspace{-2mm}
  \caption{
  Temperature dependence of the transverse Peltier coefficient $\alpha_{xy}(T)$ for an ultrathin underdoped Bi$_2$Sr$_2$CaCu$_2$O$_{8+x}$ film under various perpendicular magnetic fields, adapted from Ref.~\cite{hu2024vortex}. 
  Symbols denote experimental data, while solid lines represent self-consistent Gaussian TDGL fits, demonstrating that the thermoelectric response is governed by vortex-like superconducting fluctuations persisting well above $T_c$.
  }
  \label{fig:alpha_xy_vortex_Bi2212}
  \vspace{-2mm}
\end{figure}

As shown in Figure~\ref{fig:alpha_xy_vortex_Bi2212}, recent experiments on ultrathin underdoped Bi$_2$Sr$_2$CaCu$_2$O$_{8+x}$ reveal that the temperature and magnetic-field dependence of the transverse Peltier coefficient $\alpha_{xy}(T)$ provides a direct thermodynamic measure of the entropy carried by fluctuating vortices persisting above the superconducting transition temperature $T_c$~\cite{hu2024vortex}. 
The experimental data are quantitatively reproduced by self-consistent Gaussian TDGL calculations using a single set of representative parameters, with averaged values $T_{c0} \simeq 80.9\,\mathrm{K}$, ${B}_{c2} \simeq 161.2\,\mathrm{T}$, $k \simeq 0.9$, and $\omega \simeq 2.24\times10^{-2}$, demonstrating that superconducting fluctuations dominate the transverse thermoelectric response over a wide temperature range beyond mean-field expectations.

\begin{figure}[H]
  \centering
  \includegraphics[width=\linewidth,keepaspectratio]{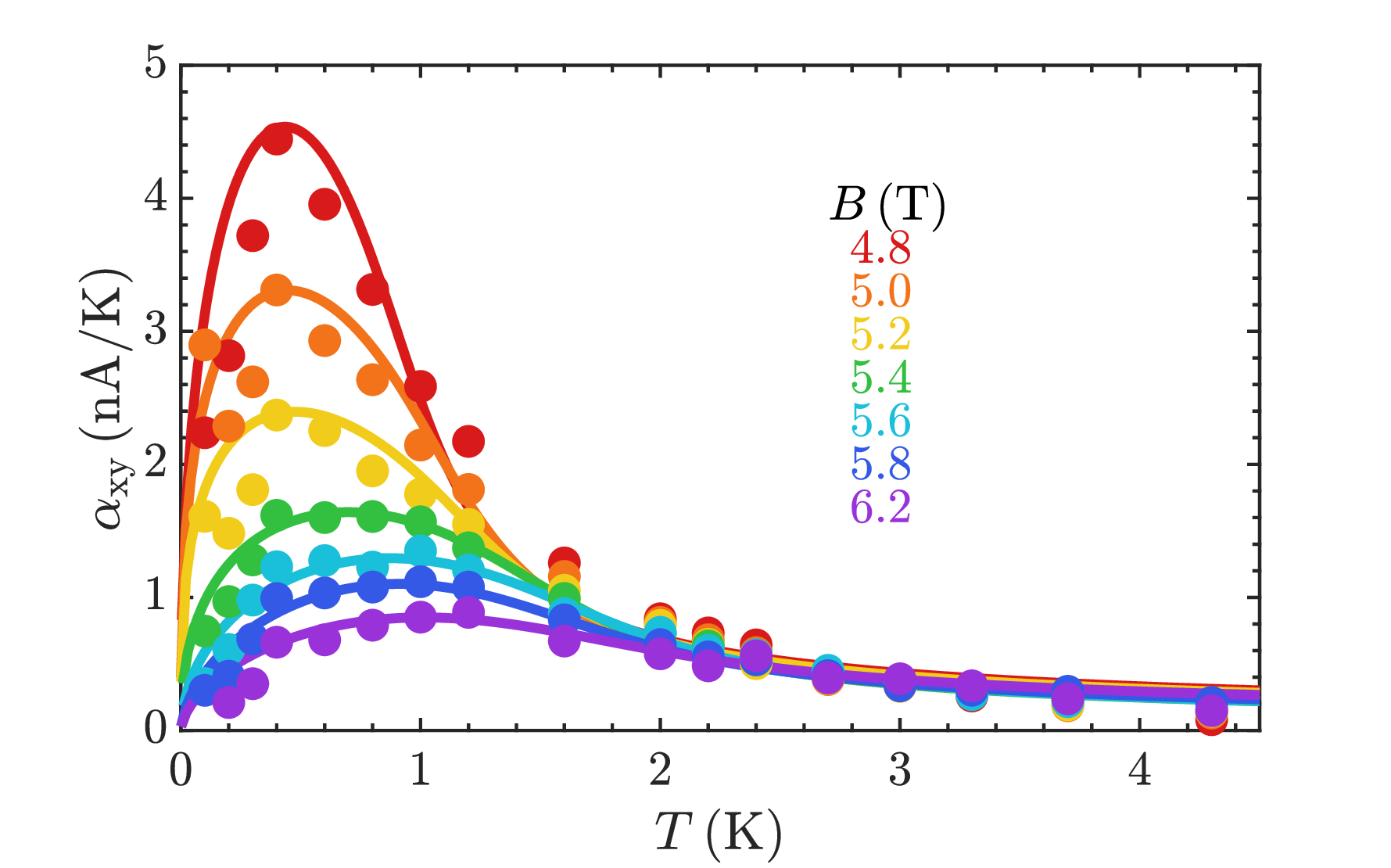}
  \vspace{-2mm}
  \caption{
  Temperature dependence of the transverse Peltier coefficient $\alpha_{xy}(T)$ for a strongly disordered superconducting thin film under various perpendicular magnetic fields, adapted from Ref.~\cite{ienaga2024broadened}. 
  Symbols represent experimental measurements, while solid curves correspond to self-consistent Gaussian TDGL fits, highlighting the persistence of fluctuation-dominated thermoelectric transport deep into the resistive state near a broadened quantum critical regime.
  }
  \label{fig:alpha_xy_disordered_QCP}
  \vspace{-2mm}
\end{figure}

A closely related phenomenology is observed in strongly disordered superconducting thin films, as reported by Ienaga \textit{et al.} in Ref.~\cite{ienaga2024broadened} and shown in Fig.~\ref{fig:alpha_xy_disordered_QCP}. 
In their experiments, the transverse Peltier coefficient $\alpha_{xy}$ remains finite deep into the resistive state and exhibits scaling behavior characteristic of fluctuation-dominated transport in the vicinity of a broadened disorder-driven quantum critical regime. 
Despite the substantial differences in microscopic pairing mechanisms and disorder strength compared with cuprate thin films, the experimental data of Ienaga \textit{et al.} are quantitatively reproduced by the Gaussian TDGL-based fluctuation theory, underscoring the universal role of superconducting fluctuations in governing transverse thermoelectric phenomena across disparate material classes.

Crucially, describing $\alpha_{xy}$ is not a decorative add-on to longitudinal transport fits. Together with the broadened resistive transitions under $B_\perp$ (Sec.~III) and the nonlinear S-shaped $I$--$V$ characteristics under voltage control (Sec.~IV), the transverse thermoelectric response completes a closed set of probes that interrogate, respectively, fluctuation-induced conductivity, nonequilibrium dynamical stability, and entropy transport. Their simultaneous quantitative description within a single mesoscopic TDGL framework with self-consistent Gaussian approximation underscores the universality and predictive power of the approximation across low-$T_c$ disordered films and high-$T_c$ cuprates, and provides a natural pathway to extend the same logic to iron-based and other reduced-dimensional superconductors.

\section{Discussion}

The results presented above establish a coherent fluctuation-based picture of transport and thermodynamics in two-dimensional superconducting films under magnetic field and electric-field drive. 
Within the time-dependent Ginzburg--Landau framework supplemented by a self-consistent Gaussian (Hartree) treatment of thermal Cooper-pair fluctuations, we have demonstrated that a single mesoscopic theory simultaneously captures three experimentally accessible and conceptually distinct phenomena: (i) the broadening of resistive transitions under perpendicular magnetic fields, (ii) intrinsic nonlinear and multivalued transport manifested as S-shaped $I$--$V$ characteristics under voltage control, and (iii) transverse thermoelectric response quantified by the Peltier coefficient $\alpha_{xy}$.

A key outcome is that fluctuation renormalization at the Gaussian--Hartree level is not restricted to a narrow vicinity of the mean-field transition temperature. 
Instead, interacting short-lived Cooper pairs generate substantial excess conductivity over extended temperature and field ranges, naturally producing smooth resistive crossovers without invoking extrinsic inhomogeneity or auxiliary transport channels. 
This mechanism is shown to be robust across a wide variety of material platforms, including atomically thin transition-metal dichalcogenides, strongly disordered low-$T_c$ films, unconventional nickelates, and ultrathin high-$T_c$ cuprates, highlighting the material-independent character of the fluctuation-controlled regime.

The nonlinear transport sector provides a particularly stringent test of the theory. 
By retaining the self-consistent feedback of superconducting fluctuations under finite electric field, the TDGL--Hartree framework predicts a nonmonotonic superconducting contribution to the current, leading to an intrinsic S-shaped instability characterized by negative differential conductivity.
Within the surface representation of the nonlinear response, this instability corresponds to a fold of the $J(E,B_\perp)$ manifold. 
The terminal points of this fold define characteristic instability scales $B_\chi$, where
\[
\left.\frac{\partial J}{\partial E}\right|_{B_\chi}=0,
\qquad
\left.\frac{\partial^{2} J}{\partial E^{2}}\right|_{B_\chi}=0,
\]
marking the disappearance of multivalued solutions and the restoration of a single-valued transport response.
Physically, $B_\chi$ therefore represent fluctuation-controlled stability boundaries rather than thermodynamic phase transitions, separating regimes with and without intrinsic nonlinear instability.

The predictive power of this construction becomes especially transparent when confronted with voltage-controlled experiments capable of stabilizing the otherwise unstable branch.
In this context, the recent resolution of continuous S-shaped $I$--$V$ curves in genuinely two-dimensional superconducting heterostructures under magnetic field provides a decisive validation of the fluctuation-driven scenario.
The ability of the present framework to reproduce both the existence of the S-shaped branch and its systematic suppression with increasing field or temperature places nonlinear transport on the same quantitative footing as linear resistance and transverse thermoelectric response.

Finally, the transverse Peltier coefficient $\alpha_{xy}$ completes this unified picture by providing a thermodynamic probe of entropy transport carried by superconducting fluctuations.
Together with resistive broadening and nonlinear instability, $\alpha_{xy}$ forms a closed and internally consistent set of observables that interrogate conductivity, dynamical stability, and entropy flow within a single mesoscopic theory.
Their simultaneous quantitative description underscores the breadth and internal consistency of the self-consistent Gaussian TDGL approach.

\section{Conclusions}

In conclusion, we have developed a unified fluctuation-based description of transport in two-dimensional superconducting films within the time-dependent Ginzburg--Landau equation supplemented by a self-consistent Gaussian (Hartree) approximation. 
This framework yields closed expressions for the renormalized resistance $R(T,B_\perp)$, the nonlinear current response $J(E,B_\perp)$, and the transverse thermoelectric coefficient $\alpha_{xy}$, enabling direct quantitative comparison with multi-field experimental datasets across distinct low-dimensional superconducting platforms.

A central outcome is the emergence of an \emph{intrinsic S-shaped} nonlinear $I$--$V$ (or $J$--$E$) characteristic generated purely by fluctuation feedback under electric-field drive. 
Within the TDGL--Hartree mechanism, the electric field self-consistently renormalizes the fluctuation spectrum, rendering the superconducting contribution to the current nonmonotonic and producing a negative-differential segment ($dI/dV<0$) with multivalued solutions under voltage control. 
This provides a concrete nonequilibrium fingerprint of fluctuation-dominated superconductivity that is absent in ordinary metallic transport. 
Moreover, the S-shaped branch is not merely reproduced qualitatively,
which mark the disappearance of multivalued transport and the restoration of a single-valued nonlinear response as temperature or magnetic field suppresses fluctuations.

Taken together, the simultaneous account of broadened resistive crossovers, the fluctuation-induced S-shaped instability in nonlinear transport, and the transverse thermoelectric response establishes the self-consistent Gaussian TDGL approach as a predictive and material-independent framework for driven superconductivity in reduced dimensions, and provides a practical route for extracting fluctuation parameters and instability scales from modern voltage-controlled experiments.

\section*{Acknowledgment}
We acknowledge Prof. Stefan Kirchner and Prof. Baruch Rosenstein for fruitful discussions.

\appendix



\bibliographystyle{unsrt} 
\bibliography{refs2}






\end{document}